\documentclass{elsarticle} 
\usepackage[outline]{contour} 
\contourlength{0.4pt}
\makeatletter
	\def\ps@pprintTitle{%
 	\let\@oddhead\@empty
	\let\@evenhead\@empty
	\def\@oddfoot{\centerline{\thepage}}%
	\let\@evenfoot\@oddfoot}
\makeatother
\usepackage[dvipsnames]{xcolor}
\usepackage[english]{babel}
\usepackage{amsfonts}
\usepackage[T1]{fontenc}
\usepackage{algorithm}
\usepackage{algpseudocode} 

\usepackage[margin=2cm]{geometry}
\usepackage{tikz}
\usepackage[utf8]{inputenc}
\usepackage{pgfplots} 
\usepackage{pgfgantt}
\usepackage{pdflscape}
\pgfplotsset{compat=newest} 
\pgfplotsset{plot coordinates/math parser=false} 
\newlength\fwidth
\newlength\fheight
\usepackage{amsmath}
\usepackage{amsthm}
\usepackage{graphicx}
\usepackage{subcaption}
\usepackage{overpic}
\usepackage{siunitx}
\usepackage[textsize=tiny]{todonotes}

\newcommand{\Ra}[1]{\color{black} {#1}}
\newcommand{\Rb}[1]{\color{black} {#1}}
\usepackage[colorlinks=true, allcolors=blue]{hyperref}

\theoremstyle{definition}

\usepackage{booktabs}

\newcommand{\real}{\mathbb{R}}
\newcommand{\bzero}{\mathbf{0}}
\newcommand{\hL}{\widehat{L}}
\newcommand{\tL}{\widehat{L}}

\newcommand{\Unl}{U_{\text{nl}}}
\newcommand{\Qm}{\ensuremath{\mathsf{Q}}}

\newcommand{\dt}{\Delta t}
\newcommand{\y}{\mathbf y}
\newcommand{\f}{\mathbf f}
\newcommand{\tf}{\mathbf{\widehat f}}
\newcommand{\q}{\mathbf q}
\newcommand{\Ir}{\mathbf{I}_r}
\newcommand{\qdot}{\mathbf{\dot{q}}}

\newcommand{\Mhat}{\mathbf{\widehat M}}
\newcommand{\Khat}{\mathbf{\widehat K}}
\newcommand{\Mtilde}{\mathbf{\widehat M}}
\newcommand{\Ktilde}{\mathbf{\widehat K}}
\newcommand{\Ctilde}{\mathbf{\widehat C}}
\newcommand{\Chat}{\mathbf{\widehat C}}

\newcommand{\K}{\mathbf K}

\newcommand{\C}{\mathbf C}

\newcommand{\Q}{\mathbf Q}
\newcommand{\qhat}{\mathbf{\widehat q}}
\newcommand{\qtilde}{\mathbf{\widehat q}}
\newcommand{\qhatdot}{\mathbf{\dot{\widehat q}}}
\newcommand{\qhatddot}{\mathbf{\ddot{\widehat q}}}
\newcommand{\qtildedot}{\mathbf{\dot{\widehat q}}}
\newcommand{\Qhat}{\mathbf{\widehat Q}}
\newcommand{\Qhatdot}{\mathbf{\dot{\widehat Q}}}
\newcommand{\Qhatddot}{\mathbf{\ddot{\widehat Q}}}

\newcommand{\V}{\mathbf{V}_r}

\newcommand{\M}{\mathbf M}

\begin{document}
\begin{frontmatter}
\title{Lagrangian operator inference enhanced with structure-preserving machine learning for nonintrusive model reduction of mechanical systems}

 		\author[affil1]{Harsh Sharma\corref{cor1}\fnref{x}}
		 		\cortext[cor1]{Corresponding author}
		\ead{hasharma@ucsd.edu}
		\fntext[x]{These authors contributed equally.}
            \author[affil2,affil3]{David A. Najera-Flores\fnref{x}}
 		\author[affil2]{Michael D. Todd}
 		\author[affil1]{Boris Kramer}

			\address[affil1]{Department of Mechanical and Aerospace Engineering, University of California San Diego, CA, United States}
		
 		\address[affil2] {Department of Structural Engineering, University of California San Diego, CA, United States}
   \address[affil3] {ATA Engineering, Inc., San Diego, CA, United States}

\begin{abstract}
    Complex mechanical systems often exhibit strongly nonlinear behavior due to the presence of nonlinearities in the energy dissipation mechanisms, material constitutive relationships, or geometric/connectivity mechanics. Numerical modeling of these systems leads to nonlinear full-order models that possess an underlying Lagrangian structure. This work proposes a Lagrangian operator inference method enhanced with structure-preserving machine learning to learn nonlinear reduced-order models (ROMs) of nonlinear mechanical systems. This two-step approach first learns the best-fit linear Lagrangian ROM via Lagrangian operator inference and then presents a structure-preserving machine learning method to learn nonlinearities in the reduced space. The proposed approach can learn a structure-preserving nonlinear ROM purely from data, unlike the existing operator inference approaches that require knowledge about the mathematical form of nonlinear terms. From a machine learning perspective, it accelerates the training of the structure-preserving neural network by providing an informed prior (i.e., the linear Lagrangian ROM structure), and it reduces the computational cost of the network training by operating on the reduced space. The method is first demonstrated on two simulated examples: a conservative nonlinear rod model and a two-dimensional nonlinear membrane with nonlinear internal damping. Finally, the method is demonstrated on an experimental dataset consisting of digital image correlation measurements taken from a lap-joint beam structure from which a predictive model is learned that captures amplitude-dependent frequency and damping characteristics accurately.  The numerical results demonstrate that the proposed approach yields generalizable nonlinear ROMs that exhibit bounded energy error, capture the nonlinear characteristics reliably, and provide accurate long-time predictions outside the training data regime. 
\end{abstract}
\end{frontmatter}
\section{Introduction}
\label{sec:introduction}
Reduced-order models (ROMs) of nonlinear mechanical models play a key role in a variety of tasks ranging from control of soft robotics~\cite{thieffry2018control,adibnazarifull} to design optimization of mechanical structures~\cite{amsallem2015design,park2013reduced} to state assessment for structural health monitoring~\cite{rosafalco2021online,taddei2018simulation}.  Nonlinear mechanical models in structural and mechanical engineering applications often possess an underlying Lagrangian structure. Deriving a Lagrangian ROM of these nonlinear mechanical models is of particular importance because the Lagrangian structure is intimately connected to physically interpretable quantities such as momentum and energy, or in case of fluid systems, vorticity. Structure-preserving model reduction of mechanical systems was introduced in~\cite{lall2003structure} where the authors showed that performing a Galerkin projection on the Euler-Lagrange equation leads to a Lagrangian ROM. The work in~\cite{carlberg2015preserving} presented a computationally efficient, structure-preserving model reduction method for parametric Lagrangian systems with higher-order nonlinearities. Both of these structure-preserving model reduction approaches are intrusive in that they require access to full-order model (FOM) operators in order to derive nonlinear ROMs via the intrusive projection. This type of information, however, is typically unavailable when working with proprietary software, complicated legacy code, or experimental data. Thus, nonintrusive methods that can learn ROMs directly from simulated or experimental data have become increasingly popular.

The operator inference framework~\cite{peherstorfer2016data} is a promising data-driven approach for learning low-dimensional models of FOMs with linear or low-order polynomial nonlinear terms. Using lifting transformations, this model reduction method has been extended to a broader class of FOMs with nonpolynomial nonlinear terms in~\cite{qian2019transform,qian2020lift,SKHW2020_learning_ROMs_combustor,khodabakhshi2022non} and to the gray-box setting in~\cite{benner2020operator} where the authors use knowledge about the nonpolynomial terms in analytic form to learn nonlinear ROMs from data. Recently, a variety of papers have embedded problem-specific structure in the operator inference framework to develop structure-preserving operator inference methods for Hamiltonian systems~\cite{sharma2022hamiltonian,gruber2023canonical} and Lagrangian mechanical systems~\cite{sharma2022preserving,filanova2023operator}. {\Ra{Similarly}} to~\cite{benner2020operator}, these structure-preserving operator inference approaches assume prior knowledge about the nonlinear terms in analytic form. Such information is typically unavailable when the underlying physics is not well known or otherwise difficult to model.    

As an alternative to learning interpretable polynomial dynamical systems, deep learning-based ROMs~\cite{hesthaven2018non,san2018neural,gao2020non,lee2020model,fresca2021comprehensive,fresca2022pod} have been proposed for nonintrusive model reduction of nonlinear PDEs. These methods exploit deep neural networks to learn both the nonlinear trial manifold and the reduced dynamics. However, these methods are not designed to preserve the underlying geometric structure and therefore may not provide accurate predictions outside the training data regime. Discovering nonlinear dynamical systems from data was first considered in~\cite{schmidt2009distilling,bongard2007automated} where the authors employed symbolic regression methods to learn the governing equations. Dictionary-based approaches based on sparse identification of nonlinear dynamical systems (SINDy)~\cite{brunton2016discovering} have also been developed for structure-preserving sparse identification of nonlinear dynamics in~\cite{kaiser2018discovering,chu2020discovering,lee2022structure}. In another research direction, the machine learning community has developed a wide variety of structure-preserving machine learning methods by endowing black-box neural networks with physics-motivated inductive biases. These structure-preserving machine learning (SpML) methods were first developed in the context of conservative dynamical systems by preserving the geometric structure related to the underlying Hamiltonian~\cite{greydanus2019hamiltonian,chen2019symplectic,jin2020sympnets,chen2021data}, Lagrangian~\cite{lutter2018deep,cranmer2020lagrangian,gupta2020structured,allen2020lagnetvip,najera2023structure}, and conservation laws~\cite{lee2021deep,van2023energy}. Building on the work in this direction, these methods have also been generalized to nonconservative systems by preserving the metriplectic structure~\cite{lee2021machine,zhang2022gfinns}. {\Ra{The authors in~\cite{toth2019hamiltonian,bertalan2019learning,saemundsson2020variational,zhong2020unsupervised,allen2020lagnetvip,qian2022trajectory,mason2023learning} have combined the aforementioned SpML methods with various extensions of autoencoders~\cite{kingma2014auto,rezende2014stochastic} to learn and predict Lagrangian/Hamiltonian dynamics from high-dimensional image datasets. Even though the SpML methods have been successful at learning nonlinear dynamics purely from data consisting of either state trajectories or high-dimensional image observations,}} a majority of these approaches are only concerned with learning {\Ra{tasks where}} the data is coming from very low-dimensional systems, e.g. $5-10$ dimensions. Therefore, these SpML methods are ill-suited for {\Ra{applications where the underlying dynamical system of interest is itself high-dimensional.}}

The main goal of this work is to develop a structure-preserving nonintrusive model reduction method that can learn ROMs strictly from observed data without assuming additional knowledge about the form of nonlinearity. The main contributions of this work are:
\begin{enumerate}
    \item We develop a Lagrangian operator inference method enhanced with structure-preserving machine learning that parametrizes the nonlinear ROM through a reduced Lagrangian and then learns both the linear and the nonlinear ROM operators in two steps. First, we learn the linear ROM operators from projections of the full-order model snapshot data via structure-preserving operator inference.  We then use structure-preserving neural networks to learn nonlinear terms in the reduced potential energy and the reduced dissipation function. 
    \item We present results for simulated data that demonstrate the proposed method's ability to provide accurate and stable predictions outside the training time interval for a conservative rod model and a nonconservative two-dimensional membrane model. 
    \item We present results for experimental data consisting of digital image correlation measurements of a lap-joint beam structure that demonstrate the applicability of the proposed approach when an underlying FOM is not available.
\end{enumerate}

The paper is structured as follows. Section~\ref{sec:background} outlines the Lagrangian mechanics formulation of nonlinear mechanical models and presents a brief review of the intrusive structure-preserving model reduction of these mechanical FOMs. Section~\ref{sec:method} presents the proposed Lagrangian operator inference method enhanced with structure-preserving machine learning for learning nonlinear Lagrangian ROMs from data. In Section~\ref{sec:numerical}, we apply the proposed method to three datasets with increasing complexity: simulated data from a conservative rod model, simulated data from a two-dimensional nonlinear membrane model with internal damping, and experimental data from a jointed structure. Section~\ref{sec:conclusions} provides concluding remarks and suggests future research directions motivated by this work.
\section{Background}
\label{sec:background}
In Section~\ref{sec:mechanics} we introduce the Lagrangian formulation of the nonlinear mechanical FOMs considered in this work. In Section~\ref{sec:intrusive} we review the construction of projection-based intrusive Lagrangian ROMs.
\subsection{Lagrangian mechanics}
\label{sec:mechanics}
Consider a nonlinear mechanical system with a configuration manifold $\Qm=\real^n$ where $n$ denotes the degrees of freedom in the FOM. We take the Lagrangian viewpoint on the mechanical system where the dynamics are described by the scalar Lagrangian function
\begin{equation}
    L(\q,\qdot)=T(\qdot) - U(\q),
\end{equation}
where $\q(t)$ is the set of generalized coordinates describing the system state,  $T(\qdot)$ is the scalar kinetic energy function, and $U(\q)$ is the scalar potential energy function. The governing equations for Lagrangian mechanical systems with nonconservative forcing $\f(\qdot)$ can be derived via the Lagrange-d'Alembert principle, see~\cite{arnol2013mathematical} for more details. This principle seeks $\q(t)$ satisfying $
        \delta \int^{t_K}_{t_0} L(\q, \qdot) \  \mathrm{d}t + \int^{t_K}_{t_0}\f(\qdot) \cdot \delta \q  \ \mathrm{d}t = 0,$
where $\delta$ represents variations vanishing at initial time $t_0$ and final time $t_K$. Using integration by parts and setting the variations at the endpoints to zero yields the forced Euler-Lagrange equations 
\begin{equation}
    \frac{\mathrm{d}}{\mathrm{d}t}\left( \frac{\partial L}{\partial \dot{\q}} \right) - \frac{\partial L}{\partial \q} = \f(\qdot).
    \label{eq:fel}
\end{equation}

For this work, we focus on FOM Lagrangians of the form 
\begin{equation}
    L(\q,\qdot)=\frac{1}{2}\qdot^\top \M\qdot - \frac{1}{2}\q^\top \K\q - \Unl(\q),
    \label{eq:fom_lagrangian}
\end{equation}
where $\M\in \real^{n \times n}$ denotes the symmetric positive-definite mass matrix, $\K \in \real^{n \times n}$ denotes the linear stiffness matrix, and $\Unl(\q)$ is the higher-order nonlinear component of the potential energy function. We note that the assumption about the quadratic form of the system's kinetic energy in~\eqref{eq:fom_lagrangian} is quite general in mechanical and structural engineering applications.  In mechanical models obtained via semi-discretization of PDEs, the linear stiffness matrix $\K$ is typically a symmetric positive-definite matrix.

The most popular approach to model damping in the context of mechanical and structural engineering applications is to assume viscous damping. Under this assumption, the nonconservative forcing in~\eqref{eq:fel} can be written as 
\begin{equation}
    \f(\qdot)=-\frac{\partial \mathcal{F}(\qdot)}{\partial \qdot},         \label{eq:fom_fgen}
\end{equation}
where $\mathcal{F}(\qdot)\geq 0$ is the scalar nonnegative dissipation function. The dissipative force $ \f(\qdot)$ is typically decomposed into linear and nonlinear components, i.e.,
\begin{equation}
    \f(\qdot)=- \C\qdot-\frac{\partial \mathcal{F}_{\text{nl}}(\qdot)}{\partial \qdot}, 
    \label{eq:fom_f}
\end{equation}
where $\C \in \real^{n \times n}$ is the linear damping matrix and $\mathcal F_{\text{nl}}(\qdot)$ is the higher-order nonlinear component of the scalar dissipation function $\mathcal{F}(\qdot)$ for modeling nonlinear damping behavior \cite{adhikari2001damping}. The linear dissipation behavior in mechanical and structural engineering applications is modeled using Rayleigh damping where the linear damping matrix is proportional to the mass and stiffness matrix. As a result, the damping matrix $\C$ also possesses a symmetric positive-definite structure. Substituting expressions for the system Lagrangian~\eqref{eq:fom_lagrangian} and the nonconservative forcing~\eqref{eq:fom_f} into the forced Euler-Lagrange equations~\eqref{eq:fel} yields the governing equations for the Lagrangian FOM
    \begin{equation}
    \M\ddot{\textbf{q}} + \C\qdot + \frac{\partial \mathcal{F}_{\text{nl}}(\qdot)}{\partial \qdot}  + \K\q+\frac{\partial \Unl(\textbf{q})}{\partial \textbf{q}}= \bzero. 
        \label{eq:fom}
\end{equation}
Variational integrators~\cite{sharma2020review} provide a principled way of deriving structure-preserving numerical integrators that respect the underlying Lagrangian structure. In this work, we use the Newmark integrator for the structure-preserving numerical integration of both the Lagrangian FOMs and the Lagrangian ROMs.
\subsection{Structure-preserving intrusive model reduction for mechanical systems}
\label{sec:intrusive}
Reduced-order models obtained via the intrusive process of Galerkin projection approximate the FOM configuration space $\Qm=\real^n$ by a reduced-order configuration space $\Qm_r=\real^r$. Using an orthonormal basis matrix $\V \in \real^{n \times r}$, the full-order model state is approximated via $\q\approx\V\qtilde$ where the reduced state is denoted with $\qhat\in \real^r$. We follow~\cite{carlberg2015preserving} to derive governing equations for the intrusive Lagrangian ROM. The intrusive reduced Lagrangian $\tL$ is defined as
\begin{equation}
\tL(\qtilde,\qtildedot):=L(\V\qtilde,\V\qtildedot)=T(\V\qtildedot) - U(\V\qtilde).
        \label{eq:Ltilde_gen}
\end{equation}
Applying the Lagrange-d'Alembert principle in the reduced setting, the resulting forced Euler-Lagrange equations in $r$ dimensions are
\begin{equation}
        \frac{\partial \tL(\qtilde,\qtildedot)}{\partial \qtilde} -\frac{\text d}{\text dt} \left( \frac{\partial \tL(\qtilde,\qtildedot)}{\partial \qtildedot} \right) + \tf(\qtildedot)  = \bzero,
    \label{eq:ELr_gen}
\end{equation}
where $\tf(\qtildedot):=\V^\top\f(\V\qtildedot)$ is the intrusive reduced nonconservative forcing.

For FOM Lagrangians of the form~\eqref{eq:fom_lagrangian}, the intrusive reduced Lagrangian is
\begin{equation}
\tL(\qtilde,\qtildedot)=\frac{1}{2}\qtildedot^\top \left(\V^\top\M\V \right)\qtildedot - \frac{1}{2}\qtilde^\top \left(\V^\top\K\V \right)\qtilde - \widehat{U}_{\text{nl}}(\qtilde),
        \label{eq:Ltilde}
\end{equation}
where $\widehat{U}_{\text{nl}}(\qtilde):=\Unl(\V\qtilde)$ is the intrusive reduced nonlinear potential energy function. The intrusive reduced nonconservative forcing $\tf(\qtildedot)$ corresponding to nonconservative forcing of the form~\eqref{eq:fom_f} is
\begin{equation}
\tf(\qtildedot)=-\left(\V^\top\C\V \right)\qtildedot-\frac{\partial \widehat{\mathcal{F}}_{\text{nl}}(\qtildedot)}{\partial \qtildedot}, 
        \label{eq:ftilde}
\end{equation}
where $\widehat{\mathcal{F}}_{\text{nl}}(\qtildedot):=\mathcal{F}_{\text{nl}}(\V\qtildedot)\geq 0$ is the nonnegative intrusive reduced dissipation function. Substituting expressions for the intrusive reduced Lagrangian~\eqref{eq:Ltilde} and the intrusive reduced nonconservative forcing~\eqref{eq:ftilde} into the reduced forced Euler-Lagrange equation leads to
\begin{equation}
    \Mtilde\mathbf{\ddot{\widehat{q}}} + \Ctilde\qtildedot + \frac{\partial \widehat{\mathcal{F}}_{\text{nl}}(\qtildedot)}{\partial \qtildedot}  + \Ktilde\qtilde+\frac{\partial \widehat{U}_{\text{nl}}(\qtilde) }{\partial \qtilde}= \bzero, 
        \label{eq:rom_intrusive}
\end{equation}
where $\Mtilde:=\V^\top\M\V$, $\Ctilde:=\V^\top\C\V$, and $\Ktilde:=\V^\top\K\V$ are the intrusive reduced operators. 

The authors in~\cite{lall2003structure,carlberg2015preserving} showed that the intrusive Lagrangian ROM~\eqref{eq:rom_intrusive} respects the symmetric positive-definite property of the system matrices and therefore, preserves the underlying Lagrangian structure. However, deriving an intrusive Lagrangian ROM requires access to FOM operators, which is often not possible when working with proprietary software or a complicated community code.  The contributions of this work are presented in the next section, where we remove the intrusive assumption and learn a nonlinear Lagrangian ROM directly from data.
\section{Lagrangian operator inference enhanced with structure-preserving machine learning}
\label{sec:method}
Section~\ref{sec:pod} describes the projection of the high-dimensional data onto a low-dimensional subspace obtained via proper orthogonal decomposition (POD). Section~\ref{sec:problem} formulates the Lagrangian structure-preserving learning problem and proposes a two-step strategy to learn a structure-preserving nonlinear ROM purely from data. The first step presented in Section~\ref{sec:lopinf} employs the Lagrangian operator inference (LOpInf) method to infer linear reduced operators. The second step presented in Section~\ref{sec:spml} introduces an SpML method that augments the linear Lagrangian ROM inferred in Section~\ref{sec:lopinf} with nonlinear terms. Section~\ref{sec:algorithm} summarizes the proposed Lagrangian operator inference enhanced with structure-preserving machine learning (LOpInf-SpML) approach in Algorithm~\ref{alg} and provides details about the offline training stage.
\subsection{Data projection via proper orthogonal decomposition}
\label{sec:pod}
We define the snapshot data matrix
    \begin{equation}
    \label{eq:snapshot}
        \Q=[\q_1,\cdots,\q_K]\in\real^{n \times K},
    \end{equation}
where $\q_1,\cdots,\q_K$ are the discrete states of the potentially high-dimensional system~\eqref{eq:fom} at time $t_1,\cdots,t_K$.  Since it is computationally infeasible to directly learn nonlinear high-dimensional models of the form~\eqref{eq:fom}, we first perform a projection of the data onto the POD subspace, in which we then learn the nonlinear ROM. 
    
We follow~\cite{holmes2012turbulence} to build the POD basis matrix $\V$ from the FOM snapshot data matrix $\mathbf{Q}=[\q_1, \cdots, \q_K]\in \real^{n \times K}$ via the singular value decomposition
\begin{equation}
    \Q=\mathbf V\pmb{\Xi} \mathbf{W}^\top,
    \label{eq:svd}
\end{equation}
where $\mathbf V \in \real^{n \times n}$, $\pmb{\Xi} \in \real^{n \times n}$, and $\mathbf{W} \in \real^{K \times n}$. The columns of the matrices $\mathbf V$ and $\mathbf{W}$ are the left and right singular vectors of $\Q$, respectively. We assume that the singular values $\xi_1 \geq \xi_2 \geq \cdots $ in $\pmb{\Xi}$ are in descending order. The POD basis of dimension $r$ is defined by the leading $r$ columns of $\mathbf V$ that correspond to the $r$ largest singular values, and the state is approximated in the linear POD subspace as $\q\approx\V\qhat$. Using this POD basis, we obtain reduced snapshot data via the projections onto $\V$ as
 \begin{equation}
     \Qhat=\V^\top\Q=[\qhat_1, \cdots, \qhat_K] \in \real^{r \times K},
          \label{eq:qhat}
 \end{equation}
 where the reduced snapshot $\qhat_k$ at time $t_k$ is $\qhat_k=\V^\top\q_k$.
 
To learn a nonlinear ROM nonintrusively from data, we also require time-derivative data at the reduced level. Here, we compute $\qhatdot$ and $\qhatddot$ from the reduced trajectory data $\qhat$ using the eighth-order central finite difference scheme 
 \begin{align}
   \label{eq:first}
     \qhatdot_k & \approx \frac{4(\qhat_{k+1}-\qhat_{k-1})}{5\dt} -\frac{(\qhat_{k+2}-\qhat_{k-2})}{5\dt} + \frac{4(\qhat_{k+3}-\qhat_{k-3})}{105\dt} - \frac{(\qhat_{k+4}-\qhat_{k-4})}{280\dt}, \\
     \qhatddot_k & \approx -\frac{205\qhat_{k}}{72\dt^2} + \frac{8(\qhat_{k+1}+\qhat_{k-1})}{5\dt^2} -\frac{(\qhat_{k+2}+\qhat_{k-2})}{5\dt^2} + \frac{8(\qhat_{k+3}+\qhat_{k-3})}{315\dt^2} - \frac{(\qhat_{k+4}+\qhat_{k-4})}{560\dt^2}.
     \label{eq:second}
 \end{align}
Using these time-derivative approximations at the reduced level, we build the snapshot matrices of the reduced first-order and second-order time-derivative data
\begin{equation}
    \Qhatdot=[\qhatdot_1, \cdots, \qhatdot_K] \in \real^{r \times K}, \quad \quad \Qhatddot=[\qhatddot_1, \cdots, \qhatddot_K] \in \real^{r \times K}.
    \label{eq:qhatdot}
\end{equation}

 \subsection{Problem formulation}
\label{sec:problem}
The main aim of the proposed LOpInf-SpML framework is to learn a predictive nonlinear ROM from the reduced trajectory and time-derivative data while also ensuring that the learned ROM (i) is a Lagrangian system; (ii) respects the symmetric positive-definite property of system matrices; and (iii) captures the conservative and dissipative behavior accurately. Towards this goal, we postulate the nonlinear ROM form in terms of its Lagrangian ingredients, i.e.,
    \begin{equation}
    \label{eq:form_general}
        \Mhat\mathbf{\ddot{\widehat{q}}} + \Chat\qhatdot + \frac{\partial \widehat{\mathcal{F}}_{\text{nl}}(\qhatdot)}{\partial \qhatdot}  + \Khat\qhat+\frac{\partial \widehat{U}_{\text{nl}}(\qhat) }{\partial \qhat}= \bzero,
\end{equation}
where $\Mhat=\Mhat^\top\succ \bzero$ is the nonintrusive reduced mass matrix, $\Chat=\Chat^\top\succ \bzero$ is the nonintrusive reduced linear damping matrix, $\Khat=\Khat^\top\succ \bzero$ is the nonintrusive reduced linear stiffness matrix, $\widehat{\mathcal{F}_{\text{nl}}} (\qhatdot)$ is the reduced nonlinear dissipation function, and $\widehat{U}_{\text{nl}}(\qhat)$ is the reduced nonlinear potential energy function. Given the reduced snapshot data $\Qhat$ and the reduced time-derivative data $\Qhatdot$ and $\Qhatddot$, the goal is to learn the reduced Lagrangian ingredients by solving
\begin{equation}
\min_{\Mhat=\Mhat^\top\succ \bzero, \Chat=\Chat^\top\succ \bzero, \Khat=\Khat^\top\succ \bzero, \widehat{\mathcal{F}}_{\text{nl}}, \widehat{U}_{\text{nl}}}\Bigg{\lVert} \Mhat \Qhatddot  + \Chat\Qhatdot + \frac{\partial \widehat{\mathcal{F}}_{\text{nl}}(\Qhatdot)}{\partial \qhatdot} + \Khat\Qhat + \frac{\partial \widehat{U}_{\text{nl}}(\Qhat) }{\partial \qhat} \Bigg{\rVert}_F,
    \label{eq:loss_general}
\end{equation}
 where the reduced matrices $\Mhat, \Chat, \Khat \in \real^{r \times r}$, the reduced nonlinear potential energy function $\widehat{U}_{\text{nl}}: \real^{r}\to \real$, and the reduced nonlinear dissipation function $\widehat{\mathcal F}_{\text{nl}}: \real^{r}\to \real$ are learned in the reduced space. However, directly training a black-box neural network to learn the nonlinear dynamics at the reduced level can be challenging from an optimization perspective, see, e.g.,~\cite{krishnapriyan2021characterizing}. Moreover, such black-box approaches typically fail to generalize outside the training data regime, as there is no guarantee that the neural network model has learned the underlying geometric structure. 
 
We therefore propose a two-step strategy {\Rb{to solve the challenging optimization problem~\eqref{eq:loss_general}.}} {\Rb {In step 1, we begin by learning}} the linear reduced stiffness matrix $\Khat$ and the linear reduced damping matrix $\Chat$ via the Lagrangian operator inference method. {\Rb{In step 2, we add more complex nonlinearities to the ROM by learning}} the reduced mass matrix $\Mhat$, the nonlinear components of the reduced potential energy function $\widehat{U}_{\text{nl}}(\qhat)$, and the nonlinear components of the reduced dissipation function $\widehat{\mathcal{F}}_{\text{nl}}(\qhatdot)$ via a structure-preserving machine learning method. To learn a nonlinear Lagrangian ROM of the form~\eqref{eq:form_general}, we postulate a nonlinear reduced Lagrangian 
      \begin{equation}
      \label{eq:Lhat}
        \hL(\qhat,\qhatdot)=\underbrace{\frac{1}{2}\qhatdot^\top \qhatdot  - \frac{1}{2}\qhat^\top \Khat \qhat}_{ =:\hL_{\text{opinf}}(\qhat,\qhatdot)}  + \underbrace{\frac{1}{2}\qhatdot^\top\Mhat_{\text{NN}}\qhatdot}_{=:\widehat{T}_{\text{NN}}(\qhatdot)} - \widehat{U}_{\text{NN}}(\qhat),
    \end{equation}
where $\hL_{\text{opinf}}(\qhat,\qhatdot)$ is the reduced Lagrangian for the linear LOpInf ROM and $\widehat T_{\text{NN}}(\qhatdot)$ and $\widehat{U}_{\text{NN}}(\qhat)$ are the neural network parametrizations of the reduced kinetic energy and nonlinear components of the reduced potential energy function, respectively.  Similarly, we postulate a nonnegative nonlinear reduced dissipation function of the form
          \begin{equation}
          \widehat{\mathcal{F}}(\qhatdot)= \widehat{\mathcal F}_{\text{opinf}}(\qhatdot) +  \widehat{\mathcal{F}}_{\text{NN}}(\qhatdot)=\frac{1}{2}\qhatdot^\top\Chat\qhatdot + \widehat{\mathcal{F}}_{\text{NN}}(\qhatdot)\geq 0,
        \label{eq:fhat}
    \end{equation}
where $\widehat{\mathcal F}_{\text{opinf}}(\qhatdot)$ is the quadratic reduced dissipation function for the linear LOpInf ROM and $\widehat{\mathcal F}_{\text{NN}}(\qhatdot)$ is the neural network parametrization of the nonlinear components of the reduced dissipation function. {\Rb{The postulated forms of the nonlinear reduced Lagrangian in~\eqref{eq:Lhat} and the nonlinear reduced dissipation function in~\eqref{eq:fhat} represent the nonlinear reduced Lagrangian ingredients as the sum of quadratic components of the linear LOpInf ROM that are learned in step 1 and the nonlinear components (including corrections to the quadratic components) in $\widehat T_{\text{NN}}(\qhatdot)$, $\widehat{U}_{\text{NN}}(\qhat)$, and $\widehat{\mathcal F}_{\text{NN}}(\qhatdot)$ that are learned in step 2. We discuss the solution of the resulting two separate minimization problems in Section~\ref{sec:lopinf} and Section~\ref{sec:spml}.}} Substituting expressions for the nonlinear reduced Lagrangian~\eqref{eq:Lhat} and the nonlinear dissipation function~\eqref{eq:fhat} into the forced Euler-Lagrange equations~\eqref{eq:fel} yields the ROM model form
    \begin{equation}
            \label{eq:rom_form_general}
            (\Ir + \Mhat_{\text{NN}} )\mathbf{\ddot{\widehat{q}}} + \Chat\qhatdot + \frac{\partial \widehat{\mathcal{F}}_{\text{NN}}(\qhatdot)}{\partial \qhatdot} + \Khat\qhat  +\frac{\partial \widehat{U}_{\text{NN}}(\qhat) }{\partial \qhat}= \bzero.
    \end{equation}
  
Since we learn the reduced potential energy function and the reduced dissipation function in two steps, we decompose the reduced mass matrix in~\eqref{eq:form_general} as $\Mhat=(\Ir + \Mhat_{\text{NN}})$, where the {\Rb{constant}} reduced mass matrix component $\Mhat_{\text{NN}}$ is learned during the training of structure-preserving neural networks in step 2. This specific decomposition of the reduced mass matrix $\Mhat$ serves two purposes: it simplifies the constrained linear least-squares problem in Section~\ref{sec:lopinf}, and it enables the learning of a {\Rb{constant}} correction term $\Mhat_{\text{NN}}$ in the reduced mass matrix $\Mhat$ that is compatible with the nonlinear components $\widehat U_{\text{NN}}$ and $\widehat{\mathcal F}_{\text{NN}}$  learned in step 2 in Section~\ref{sec:spml}.
\begin{figure}[h]%
\centering
\includegraphics[width=\textwidth]{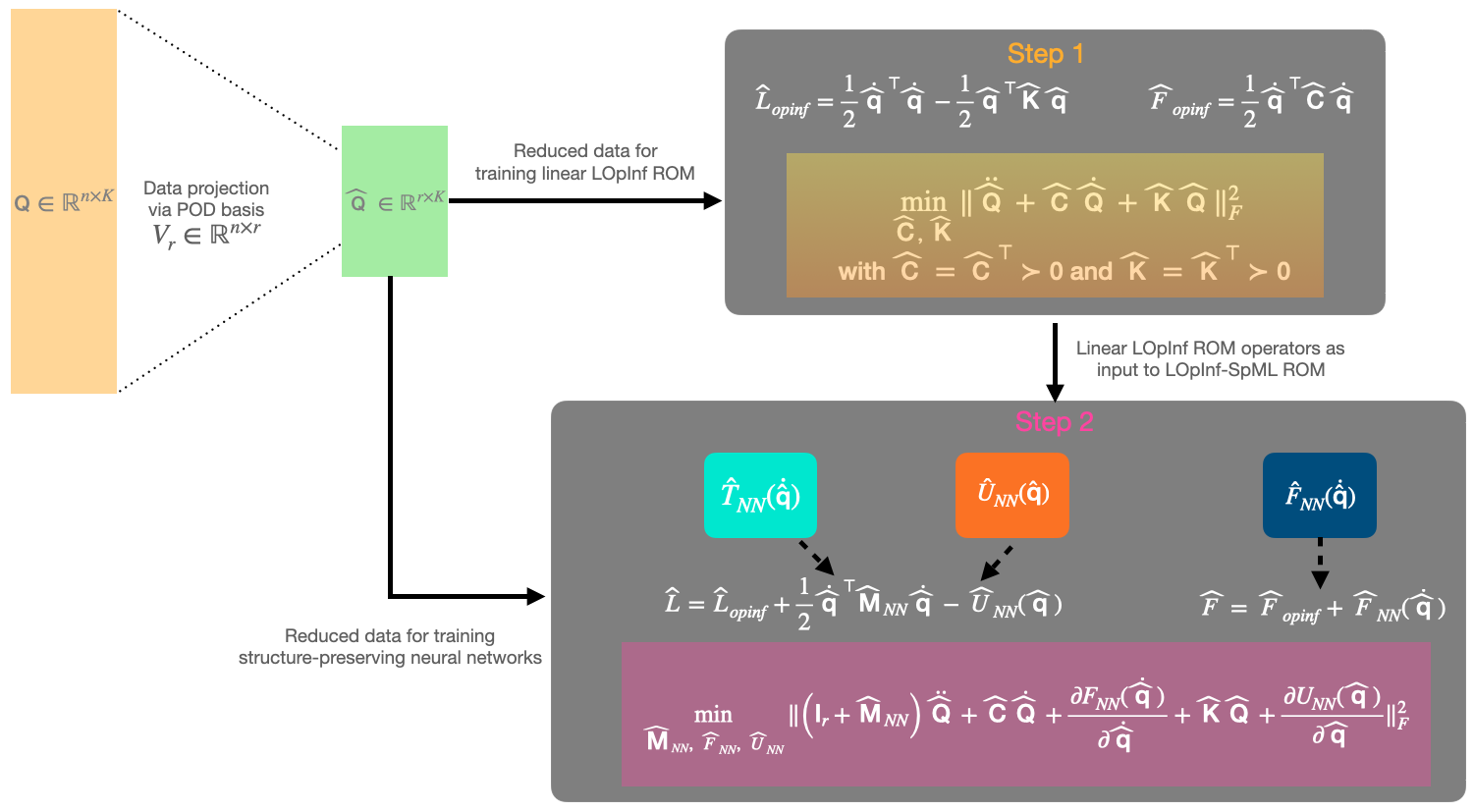}
  \caption{Architecture for the LOpInf-SpML method. Starting from the FOM snapshot data $\Q$ in~\eqref{eq:snapshot}, the reduced snapshot data $\Qhat$ in~\eqref{eq:qhat} is obtained via projections onto the POD basis matrix $\V$. Step 1 learns the linear LOpInf ROM~\eqref{eq:lopinf_rom} from the reduced snapshot data and then step 2 uses the linear LOpInf ROM as a prior to learn the nonlinear reduced operators in the LOpInf-SpML ROM~\eqref{eq:rom_form_general} from the reduced snapshot data.}
  \label{fig:schematic}
\end{figure}
\subsection{Step 1: Lagrangian operator inference method to learn the linear reduced operators}
\label{sec:lopinf}
To learn the linear parts of the Lagrangian ROM, we use the Lagrangian operator inference method~\cite{sharma2022preserving} which fits linear reduced operators to the reduced data in a structure-preserving way. For this portion of the learning process, we learn the quadratic components of the Lagrangian as
\begin{equation}
    \widehat{L}_{\text{opinf}}(\qhat,\qhatdot)=\frac{1}{2}\qhatdot^\top\qhatdot - \frac{1}{2}\qhat^\top\Khat\qhat,
\end{equation}
where $\Khat\in \real^{r \times r}$ is the symmetric positive-definite reduced stiffness matrix. Similarly, we postulate a quadratic dissipation function
\begin{equation}
    \widehat{\mathcal F}_{\text{opinf}}(\qhatdot)=\frac{1}{2}\qhatdot^\top\Chat\qhatdot,
\end{equation}
where $\Chat \in \real^{r \times r}$ is the symmetric positive-definite reduced damping matrix. 

Based on the assumed model form for the dissipation function $\widehat{\mathcal F}_{\text{opinf}}(\qhatdot)$ and the reduced Lagrangian  $\widehat{L}_{\text{opinf}}(\qhat, \qhatdot)$ in step~1, we derive the governing equations for the linear LOpInf ROM via the forced Euler-Lagrange equations
\begin{equation}
    \qhatddot + \Chat\qhatdot + \Khat\qhat=\bzero.
    \label{eq:lopinf_rom}
\end{equation}
We obtain the linear symmetric positive definite reduced operators $\Chat \in \real^{r \times r}$ and $\Khat \in \real^{r \times r}$ by solving
\begin{equation}
\min _{\substack{\Khat=\Khat^\top \succ \bzero, \Chat=\Chat^\top \succ \bzero}}
\bigg{\lVert} \Qhatddot + \Chat\Qhatdot + \Khat\Qhat \bigg{\rVert}_{F}.
\label{eq:lopinf}
\end{equation}
The symmetric positive-definite constraints on the linear reduced operators ensure that the linear LOpInf ROM~\eqref{eq:lopinf_rom} preserves the underlying Lagrangian structure. 

Since the snapshot data $\Q$ from~\eqref{eq:snapshot} is obtained (simulated or experimentally measured) from a nonlinear mechanical system, the linear LOpInf ROM in~\eqref{eq:lopinf_rom} may be inaccurate because it does not capture the underlying nonlinear behavior. Step 2 in the proposed learning approach exploits the expressive power of neural networks to augment the linear LOpInf ROM with missing nonlinear ROM operators.
\subsection{Step 2: Structure-preserving machine learning for nonlinear reduced operators}
\label{sec:spml}
In this section, we present a structure-preserving machine learning architecture to learn the nonlinear components of the reduced potential energy and the reduced dissipation function. {\Ra{For learning these nonlinear components, we choose a nonlinear parametrization that strikes a balance between polynomial terms based on domain-specific knowledge and neural networks that can capture more complex nonlinearities.  Polynomial-augmented multilayer perceptrons (MLPs)~\cite{wang2023neural,najera2023structure} provide an \textit{expressive} parametrization where the polynomial terms allow us to exactly capture the effects of  low-degree polynomial (but not purely quadratic) energy terms commonly observed in mechanical systems, and the MLPs allow us to learn arbitrary nonlinearities that can not be captured by polynomial terms.}} We parametrize the nonlinear components of the reduced potential energy as

    \begin{equation}
        \widehat{U}_{\text{NN}}(\qhat;  \boldsymbol{\alpha}, \boldsymbol{\lambda}, \boldsymbol{\theta}_{\widehat{U}_{\text{NN}}})=\sum_{i_1,i_2,\cdots,i_r}^{i_1 + i_2 + \cdots + i_r=P_1}\alpha_{i_1,i_2,\cdots,i_r}\hat{q}_1^{i_1}\hat{q}_2^{i_2}\cdots\hat{q}_r^{i_r} + \sum_{i}^N\lambda_i\mathcal{U}^{(i)}(\qhat),
    \label{eq:Unn}
    \end{equation}
    where $P_1$ is the degree of the polynomial term, $\boldsymbol{\alpha}$ contains all the unknown $\alpha_{i_1,i_2,\cdots,i_r}$ coefficients, $\boldsymbol{\lambda}$ contains all the unknown $\lambda_i$ coefficients, the function $\mathcal{U}^{(i)}(\qhat)$ represents an MLP, and $\boldsymbol{\theta}_{\widehat{U}_{\text{NN}}}$ contains all the neural network parameters used to define the MLPs in~\eqref{eq:Unn}.  {\Rb{The quadratic terms in $\widehat{U}_{\text{NN}}(\qhat)$, i.e., $P_1=2$ with $i_1 + \cdots + i_r=2$, may be interpreted as providing a correction to the linear reduced stiffness matrix $\widehat{\textbf K}$ learned in step 1.}} We parametrize the nonlinear components of the reduced dissipation function as 
    \begin{equation} \widehat{\mathcal{F}}_{\text{NN}}(\qhatdot; \boldsymbol{\beta}, \boldsymbol{\gamma}, \boldsymbol{\theta}_{\widehat{\mathcal F}_{\text{NN}}} )=\sum_{i_1,i_2,\cdots,i_r}^{i_1 + i_2 + \cdots + i_r=P_2}\beta_{i_1,i_2,\cdots,i_r}\dot{\hat{q}}_1^{i_1}\dot{\hat{q}}_2^{i_2}\cdots\dot{\hat{q}}_r^{i_r} + \sum_{i}^N\gamma_i\mathcal{F}^{(i)}(\qhatdot),
    \label{eq:Fnn}
    \end{equation}
      where $P_2$ is the degree of the polynomial term, $\boldsymbol{\beta}$ contains all the unknown $\beta_{i_1,i_2,\cdots,i_r}$ coefficients, $\boldsymbol{\gamma}$ contains all the unknown $\gamma_i$ coefficients, the function $\mathcal{F}^{(i)}(\qhatdot)$ represents an MLP, and $\boldsymbol{\theta}_{\widehat{\mathcal F}_{\text{NN}}}$ contains all the neural network parameters used to define the MLPs in~\eqref{eq:Fnn}. {\Rb{The quadratic terms in $\widehat{\mathcal F}_{\text{NN}}(\qhatdot)$, i.e., $P_2=2$ with $i_1 + \cdots + i_r=2$, may be interpreted as providing a correction to the linear reduced damping matrix $\widehat{\textbf C}$ learned in step 1.}}  In this work, we use polynomial terms up to degree $P_1=P_2=4$ in both  $\widehat{U}_{\text{NN}}(\qhat)$ and $ \widehat{\mathcal F}_{\text{NN}}(\qhatdot)$. Finally, we parametrize the reduced kinetic energy term as
    \begin{equation}
        \widehat{T}_{\text{NN}}(\qhatdot;\boldsymbol{\zeta})=\sum_{i_1,i_2,\cdots,i_r}^{i_1 + i_2 + \cdots + i_r=2}\zeta_{i_1,i_2,\cdots,i_r}\dot{\hat{q}}_1^{i_1}\dot{\hat{q}}_2^{i_2}\cdots\dot{\hat{q}}_r^{i_r},
    \label{eq:Tnn}
    \end{equation}
where we only use polynomial terms up to degree two in $ \widehat{T}_{\text{NN}}(\qhatdot)$ and $\boldsymbol{\zeta}$ contains all the unknown $\zeta_{i_1,i_2,\cdots,i_r}$ coefficients. This quadratic form for the reduced kinetic energy term $\widehat{T}_{\text{NN}}(\qhatdot)$  leads to a constant, symmetric reduced mass matrix $\Mhat_{\text{NN}}$ which is obtained by computing the Hessian of $\widehat{T}_{\text{NN}}(\qhatdot)$ as
    \begin{equation}
    [\Mhat_{\text{NN}}]_{ij} = \frac{\partial ^2 \widehat{T}_{\text{NN}}\left(\qhatdot\right)}{\partial \dot{\widehat{q}}_i \partial \dot{\widehat{q}}_j},
    \label{eq:Mnn_new}
    \end{equation}
where $ [\Mhat_{\text{NN}}]_{ij}$ denotes the $i,j$th element of the $\Mhat_{\text{NN}}$ matrix for $i,j=1,\ldots,r$.

We consider a {\Ra{unsupervised}} learning setup and learn the LOpInf-SpML ROM~\eqref{eq:rom_form_general} by solving
\begin{equation}
\label{eq:loss_min}
\min_{\boldsymbol{\zeta}, \boldsymbol{\alpha}, \boldsymbol{\beta}, \boldsymbol{\lambda}, \boldsymbol{\gamma}, \boldsymbol{\theta}_{\widehat{U}_{\text{NN}}}, \boldsymbol{\theta}_{\widehat{\mathcal F}_{\text{NN}}} } \mathcal J(\boldsymbol{\zeta}, \boldsymbol{\alpha}, \boldsymbol{\beta}, \boldsymbol{\lambda}, \boldsymbol{\gamma}, \boldsymbol{\theta}_{\widehat{U}_{\text{NN}}}, \boldsymbol{\theta}_{\widehat{\mathcal F}_{\text{NN}}} ) \qquad  \text{such that}\qquad  \frac{1}{2}\qhatdot^\top(\Ir+\Mhat_{\text{NN}})\qhatdot > 0, \  \widehat{\mathcal F}(\qhatdot) \geq 0,
\end{equation}
where the loss function $\mathcal{J}(\boldsymbol{\zeta}, \boldsymbol{\alpha}, \boldsymbol{\beta}, \boldsymbol{\lambda}, \boldsymbol{\gamma}, \boldsymbol{\theta}_{\widehat{U}_{\text{NN}}}, \boldsymbol{\theta}_{\widehat{\mathcal F}_{\text{NN}}} )$ is based on the residual of the forced Euler-Lagrange equations at the reduced level, specifically
    \begin{equation}
    \label{eq:loss}
       \mathcal{J}(\boldsymbol{\zeta}, \boldsymbol{\alpha}, \boldsymbol{\beta}, \boldsymbol{\lambda}, \boldsymbol{\gamma}, \boldsymbol{\theta}_{\widehat{U}_{\text{NN}}}, \boldsymbol{\theta}_{\widehat{\mathcal F}_{\text{NN}}} ) = \Bigg{\lVert}  (\Ir+\Mhat_{\text{NN}}) \Qhatddot + \Chat\Qhatdot + \frac{\partial \widehat{\mathcal{F}}_{\text{NN}}(\Qhatdot; \boldsymbol{\beta}, \boldsymbol{\gamma}, \boldsymbol{\theta}_{\widehat{\mathcal F}_{\text{NN}}} )}{\partial \qhatdot}  + \Khat\Qhat+\frac{\partial \widehat{U}_{\text{NN}}(\Qhat;  \boldsymbol{\alpha}, \boldsymbol{\lambda}, \boldsymbol{\theta}_{\widehat{U}_{\text{NN}}}) }{\partial \qhat}  \Bigg{\rVert}_F,
    \end{equation}
wherein the linear symmetric positive definite reduced operators $\Khat$ and $\Chat$ are learned in step 1. The reduced kinetic energy term $\widehat{T}_{\text{NN}}(\qhatdot)$, the reduced nonlinear potential energy term $\widehat{U}_{\text{NN}}(\qhat)$, and the reduced nonlinear dissipation function $\widehat{\mathcal F}_{\text{NN}}(\qhatdot)$ learned through the SpML method in step 2 can be interpreted as perturbations to the reduced Lagrangian ingredients in step 1.  The proposed LOpInf-SpML ROMs provide a flexible learning approach that combines the Lagrangian operator inference method~\cite{sharma2022preserving} with the structure-preserving neural networks.
\subsection{Complete training algorithm}
\label{sec:algorithm}
The schematic in Figure~\ref{fig:schematic} provides a visual explanation of the two-step learning strategy whereas Algorithm~\ref{alg} formally summarizes the proposed method which learns the LOpInf-SpML ROM~\eqref{eq:rom_form_general} in two separate stages. After projecting the high-dimensional data onto a low-dimensional subspace of appropriate dimension, we first learn the linear reduced operators by solving the constrained optimization problem~\eqref{eq:lopinf}. The constrained linear least-squares problems for inferring the linear reduced operators in the numerical studies are solved using semidefinite programming mode provided by the CVX package~\cite{diamond2016cvxpy}. To preserve the Lagrangian structure in the linear ROM, we declare $\Khat$ and $\Chat$ as symmetric matrix variables and impose $\Khat - 10^{-8}\cdot \Ir \succeq \bzero$ and $\Chat - 10^{-8}\cdot \Ir \succeq \bzero$. These constraints ensure that the inferred reduced operators are symmetric positive-definite matrices.

In the SpML training procedure, the projected snapshot data is split into $\Qhat = [\Qhat_{\text{train}}, \Qhat_{\text{val}}]$, where the training dataset $\Qhat_{\text{train}}$ is used for training the model parameters and the validation dataset $\Qhat_{\text{val}}$ is used for assessing the model's performance. The training data is divided into minibatches of size $N_{\text{mb}}$, which determines the number of samples that are propagated through the neural network before the weights are updated. We use a minibatch size of $N_{\text{mb}}=250$ for all the examples in Section~\ref{sec:numerical}. To preserve the Lagrangian structure in the LOpInf-SpML ROM~\eqref{eq:rom_form_general}, we numerically solve the constrained optimization problem in~\eqref{eq:loss_min} via
\begin{equation}
\label{eq:loss_min}
\min_{\boldsymbol{\zeta}, \boldsymbol{\alpha}, \boldsymbol{\beta}, \boldsymbol{\lambda}, \boldsymbol{\gamma}, \boldsymbol{\theta}_{\widehat{U}_{\text{NN}}}, \boldsymbol{\theta}_{\widehat{\mathcal F}_{\text{NN}}} } \mathcal J(\boldsymbol{\zeta}, \boldsymbol{\alpha}, \boldsymbol{\beta}, \boldsymbol{\lambda}, \boldsymbol{\gamma}, \boldsymbol{\theta}_{\widehat{U}_{\text{NN}}}, \boldsymbol{\theta}_{\widehat{\mathcal F}_{\text{NN}}} )  +   \mathcal{J}_+(\boldsymbol{\zeta}, \boldsymbol{\alpha}, \boldsymbol{\beta}, \boldsymbol{\lambda}, \boldsymbol{\gamma}, \boldsymbol{\theta}_{\widehat{U}_{\text{NN}}}, \boldsymbol{\theta}_{\widehat{\mathcal F}_{\text{NN}}} ),  
\end{equation}
 where the constraints on the total reduced kinetic energy and the reduced dissipation function are imposed through an added penalty function
    \begin{equation}
    \label{eq:loss_pos}
       \mathcal{J}_+(\boldsymbol{\zeta}, \boldsymbol{\alpha}, \boldsymbol{\beta}, \boldsymbol{\lambda}, \boldsymbol{\gamma}, \boldsymbol{\theta}_{\widehat{U}_{\text{NN}}}, \boldsymbol{\theta}_{\widehat{\mathcal F}_{\text{NN}}} ) = \Bigg{\lVert} \text{ReLU}\left(-1\left(\frac{1}{2}\Qhatdot^\top(\Ir+\Mhat_{\text{NN}})\Qhatdot\right) \right) +  \text{ReLU}\left(-1\left(\widehat{\mathcal{F}}(\Qhatdot) \right) \right) \Bigg{\rVert}_F,
    \end{equation}
    where $\text{ReLU}$ represents the rectified linear unit function which is nonzero only for positive argument values.

The neural network parametrizations in this work use the Swish activation function~\cite{ramachandran2017searching} for all the hidden layers and the linear activation function for the final layer.  The network parameters {\Rb{and the unknown coefficients in \eqref{eq:Unn} and \eqref{eq:Fnn}}} are collected in a vector $\boldsymbol{\theta} \in \real^{N_{\text{param}}}$ where $N_{\text{param}}$ is the total number of {\Rb{unknown}} parameters. The weights for each layer in {\Rb{the neural network parametrizations}} are randomly initialized using the Glorot initialization algorithm \cite{glorot2010init}. The loss function~\eqref{eq:loss} is minimized using the ADAM optimizer~\cite{kingma2015adam}, which is a widely used stochastic optimization method in the machine learning community. During the optimization process, weights inside the neural network functions along with the coefficients in \eqref{eq:Unn} and \eqref{eq:Fnn} are optimized simultaneously. For the ADAM optimizer, we use an initial learning rate $\eta=10^{-4}$ and exponential decay parameter values $\beta_1=0.9$ and $\beta_2=0.999$. 

\begin{algorithm}

\caption{Lagrangian operator inference enhanced with structure-preserving machine learning (LOpInf-SpML)}
\begin{algorithmic}[1]
\Require Full-order snapshot data $\Q\in \real^{n \times K}$  and reduced dimension $r$. Additional hyperparameters: size of training set $T_{\text{train}}$ and validation set $T_{\text{val}}$, initial learning rate $\eta$, batch size $N_b$, maximum number of epochs $N_{\text{epochs}}$, and number of minibatches $N_{\text{mb}} = (T_{\text{train}} - T_{\text{val}})/N_b$.
\Ensure Reduced linear operators $\Mhat_{\text{NN}},\Chat,\Khat \in \real^{r \times r}$, neural network for reduced nonlinear potential energy $\widehat{U}_{\text{NN}}(\qhat)$, and neural network for reduced nonlinear dissipation function $\widehat{\mathcal{F}}_{\text{NN}}(\qhatdot)$.
\State Build basis matrix $\V \in \real^{n \times r}$ from SVD of $\Q \in \real^{n \times K}$~\eqref{eq:snapshot}.
\State Project to obtain reduced state data $\Qhat \in \real^{r \times K}$ \eqref{eq:qhat}.
    \State Compute  reduced time-derivative data $\Qhatdot,\Qhatddot \in \real^{r \times K}$ as in \eqref{eq:qhatdot} from the projected data $\Qhat$ using an appropriate finite difference scheme (e.g., \eqref{eq:first}-\eqref{eq:second}).
    \State Solve constrained linear least-squares problem~\eqref{eq:lopinf} to nonintrusively infer reduced operators $\Khat$ and $\Chat$. 
    \State Split projected snapshot data into $\Qhat = [\Qhat_{\text{train}}, \Qhat_{\text{val}}]$, similarly for $\Qhatdot,\Qhatddot$.
    \State Randomly initialize weights $\boldsymbol{\theta}^0$.
    \State Initialize epoch counter $n_e = 0$.
    \While{$n_e < N_{\text{epochs}}$} 
        \For{$i = 1:N_{mb}$}
            \State Build a minibatch from the training set, $\Qhat_{\text{batch}} \subseteq  \Qhat_{\text{train}}$, similarly for $\Qhatdot_{\text{batch}},\Qhatddot_{\text{batch}}$.
            \State Evaluate $\widehat{U}_{\text{NN}}(\Qhat_{\text{batch}})$~\eqref{eq:Unn}, $\widehat{T}_{\text{NN}}(\Qhatdot_{\text{batch}})$~\eqref{eq:Tnn}, and $\widehat{\mathcal{F}}_{\text{NN}}(\Qhatdot_{\text{batch}})$~\eqref{eq:Fnn}, which are parameterized by $\boldsymbol{\theta}^{N_{\text{mb}}n_e + i}$.
            \State Evaluate loss function from residual as shown in \eqref{eq:loss}.
            \State Accumulate loss and compute gradient of loss function $\nabla_{\boldsymbol{\theta}}\mathcal{J}$.
            \State Update weights $\boldsymbol{\theta}^{N_{mb}n_e + i + 1} = \text{ADAM}(\eta, \nabla_{\boldsymbol{\theta}}\mathcal{J}, \theta^{N_{\text{mb}}n_e + i})$. 
        \EndFor
        \State Repeat \textbf{for loop} on $\Qhat_{\text{val}}, \Qhatdot_{\text{val}}, \Qhatddot_{\text{val}}$ with updated weights.
        \State Accumulate loss on  $\Qhat_{\text{val}}, \Qhatdot_{\text{val}}, \Qhatddot_{\text{val}}$ to evaluate network convergence.
        \State $n_e = n_e + 1$.
    \EndWhile 
\end{algorithmic}
\label{alg}
\end{algorithm}
\section{Numerical results}
\label{sec:numerical}
In this section, we apply the proposed LOpInf-SpML method to three nonlinear mechanical examples with increasing complexity. In Section~\ref{sec: rod} we apply the proposed method to a one-dimensional conservative rod model with isolated elastic nonlinearities to investigate its performance for conservative nonlinear mechanical systems.  In Section~\ref{sec: membrane} we study a two-dimensional geometrically nonlinear membrane model with internal damping to show the effectiveness of the proposed approach for nonconservative mechanical systems. Finally, in Section~\ref{sec: joint} we consider a dataset of experimental measurements of a lap-joint beam structure to demonstrate the proposed method's ability to learn structure-preserving surrogate models from experimental data. 
\subsection{Conservative rod model with spatially isolated elastic nonlinearities}
\label{sec: rod}
\begin{figure}[h]%
\centering

\usetikzlibrary{patterns,snakes}
\usetikzlibrary{arrows.meta} 

\colorlet{xcol}{blue!70!black}
\colorlet{darkblue}{blue!40!black}
\colorlet{myred}{red!65!black}
\tikzstyle{mydashed}=[xcol,dashed,line width=0.25,dash pattern=on 2.2pt off 2.2pt]
\tikzstyle{axis}=[->,thick] 
\tikzstyle{ell}=[{Latex[length=3.3,width=2.2]}-{Latex[length=3.3,width=2.2]},line width=0.3]
\tikzstyle{dx}=[-{Latex[length=3.3,width=2.2]},darkblue,line width=0.3]
\tikzstyle{ground}=[preaction={fill,top color=black!10,bottom color=black!5,shading angle=20},
                    fill,pattern=north east lines,draw=none,minimum width=0.3,minimum height=0.6]
\tikzstyle{mass}=[line width=0.6,red!30!black,fill=red!40!black!10,rounded corners=1,
                  top color=red!40!black!20,bottom color=red!40!black!10,shading angle=20]
\tikzstyle{spring}=[line width=0.8,blue!7!black!80,snake=coil,segment amplitude=5,segment length=5,line cap=round]
\tikzset{>=latex} 
\tikzstyle{force}=[->,myred,very thick,line cap=round]
\def\tick#1#2{\draw[thick] (#1)++(#2:0.1) --++ (#2-180:0.2)}


\begin{tikzpicture}
  \def\H{1}    
  \def\T{0.3}  
  \def\W{2.6}  
  \def\D{0.25} 
  \def\h{0.6}  
  \def\w{0.7}  
  \def\x{1}  
   \def\y{2.7}  
   \def\xx{4.4}  
   \def\yy{6.1}  
    \def\xxx{7.8}  
   \def\yyy{9.5}  
       \def\xxxx{11.2}  
   \def\yyyy{12.9}  
       \def\xxxxx{14.6}  
   \def\yyyyy{16.3}  
  \draw[spring] (0,\h/2) --++ (\x,0);
   \draw[spring] (\x + \w,\h/2) --++ (1,0);
   \draw[spring] (\y + \w,\h/2) --++ (1,0);
   \draw[spring] (\xx + \w,\h/2) --++ (1,0);
      \draw[spring] (\yy + \w,\h/2) --++ (1,0);
   \draw[spring] (\xxx + \w,\h/2) --++ (1,0);
         \draw[spring] (\yyy + \w,\h/2) --++ (1,0);
   \draw[spring] (\xxxx + \w,\h/2) --++ (1,0);
            \draw[spring] (\yyyy + \w,\h/2) --++ (1,0);
   \draw[spring] (\xxxxx + \w,\h/2) --++ (1,0);
  \draw (0,\H) --(0,0);
   \draw (\xxxxx + \w + 1,\H) --(\xxxxx + \w + 1,0);
  \draw[mass] (\x,0) rectangle++ (\w,\h) node[midway] {$m$};
   \draw[mass] (\y,0) rectangle++ (\w,\h) node[midway] {$m$};
     \draw[mass] (\xx,0) rectangle++ (\w,\h) node[midway] {$m$};
   \draw[mass] (\yy,0) rectangle++ (\w,\h) node[midway] {$m$};
        \draw[mass] (\xxx,0) rectangle++ (\w,\h) node[midway] {$m$};
   \draw[mass] (\yyy,0) rectangle++ (\w,\h) node[midway] {$m$};
           \draw[mass] (\xxxx,0) rectangle++ (\w,\h) node[midway] {$m$};
   \draw[mass] (\yyyy,0) rectangle++ (\w,\h) node[midway] {$m$};
           \draw[mass] (\xxxxx,0) rectangle++ (\w,\h) node[midway] {$m$};
    \draw[->]        (\xx+1.15,-0.25)   -- (\xx +1.5,1);
    \draw[->]        (\yy+1.15,-0.25)   -- (\yy +1.5,1);
    \draw[->]        (\xxx+1.15,-0.25)   -- (\xxx +1.5,1);
    \draw[dashed] (\xx+\w/2,0) --(\xx + \w/2,-0.5);
    \draw[dashed] (\yyy+\w/2,0) --(\yyy + \w/2,-0.5);
    \node at (\xx + \w/2 + 0.1,-0.7) {$s_1$};
    \node at (\yyy + \w/2 + 0.1,-0.7) {$s_2$};
\end{tikzpicture}

  \caption{Conservative rod model. The nonlinearities are localized within the region $s \in (s_{1}, s_{2})$, while the remainder of the rod is linear.}
  \label{fig:rod}
\end{figure}
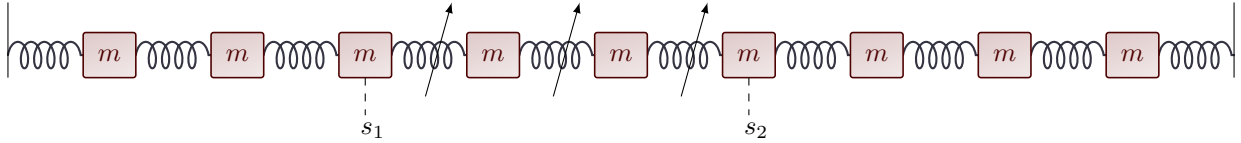

We consider the conservative structural dynamics of a one-dimensional rod with spatially isolated elastic nonlinearities. The rod is modeled with a spring-mass model consisting of a chain of masses connected with spring elements. A schematic of this nonlinear mass-spring model is shown in Figure~\ref{fig:rod}. 
\subsubsection{FOM implementation}
For this numerical experiment, we consider a rod of length $\ell=1$ and discretize it with a chain of $n=64$ mass elements with equal mass $m=1.56\times 10^{-2}$ which leads to a state vector of dimension $\q\in \real^{64}$. The masses are connected with spring elements with a constant linear stiffness $\kappa=65.0$. The region with isolated nonlinearities, i.e. $s\in(0.25,0.35)$, is modeled with additional cubic spring elements with coefficient $\varrho=2.62 \times 10^5$. As a result, the interaction force $f_{i}(q_i,q_{i+1})$ between two adjacent masses is
    \begin{equation*}
        f_{i}(q_i,q_{i+1})= \begin{cases}
        \kappa(q_{i+1} - q_i) + \varrho(q_{i+1} - q_i)^3  & \text{if } i \in \{22,23,24,25,26,27,28\}\\
         \kappa(q_{i+1} - q_i) & \text{otherwise} 
    \end{cases},
    \end{equation*}
    where $q_i$ represents the displacement of the $i$th element. For numerical comparison with intrusive Lagrangian ROMs, we provide details about the Lagrangian FOM for this example. The FOM Lagrangian for this conservative mechanical system can be written as
    \begin{equation}
    \label{eq:beam_lagrangian}
        L(\q,\qdot)=\frac{1}{2}\qdot^\top \M\qdot -\frac{1}{2}\q^\top \K\q - \sum_{i=22}^{28} \frac{\varrho}{4} \left(q_{i+1}-q_i\right)^4,
    \end{equation}
    where  $\M\in \real^{64 \times 64}$ and  $\K\in \real^{64 \times 64}$ are the symmetric positive-definite system matrices. The governing equations for the Lagrangian FOM are
    \begin{equation}
        \M\ddot{\q} - \frac{\partial L(\q,\qdot)}{\partial \q}=\bzero.         \label{eq:rod_fom}
    \end{equation}
We specify the initial condition in terms of modal velocities
$ \q(0)=\bzero$ and $\qdot(0)=\boldsymbol{\Phi} \begin{bmatrix} \nu_1,\cdots, \nu_{64} \end{bmatrix}^\top$,
where $\boldsymbol{\Phi}\in \real^{64 \times 64}$ is the modal transformation matrix based on the linear modal analysis and $\nu_i$ values are the initial modal velocity coefficients. 
\subsubsection{Learning setup}
In this study, we consider a training initial condition with nonzero initial modal velocity for the first three modes
\begin{equation}
        \nu_1=1.0 \times 10^{-1},\quad  \nu_2=2.5 \times 10^{-2},\quad  \nu_3=5.0 \times 10^{-2},  \quad \nu_i=0 \ \text{for} \ 4\leq i \leq 64.
        \label{eq:ic}
\end{equation}
We generate the simulated snapshot data by numerically integrating the conservative FOM~\eqref{eq:rod_fom} for total time $T=16$ using a Newmark integrator with a fixed time step of $\dt=10^{-3}$. We use data from $t=0$ to $t=7.5$ as the training data, data from $t=7.5$ to $t=8$ as the validation data, and data from $t=8$ to $T=16$ as the test data. We consider ROMs of size $r=4$, $r=6$, and $r=8$ to show the increase in the accuracy of the learned ROMs with an increase in the reduced dimension.

Since the high-dimensional system is a conservative mechanical system, we learn a nonlinear ROM of the form
    \begin{equation}
        \qhatddot + \Khat\qhat + \frac{\partial \widehat{U}_{\text{NN}}(\qhat)}{\partial \qhat} = \bzero,
        \label{eq:rom_rod}
    \end{equation}
where $\Khat$ is the reduced linear stiffness matrix and $\widehat{U}_{\text{NN}}(\qhat)$ is the neural network component of the reduced potential energy. The specific choice for the reduced mass matrix, i.e., $\Mhat=\Ir$, in~\eqref{eq:rom_rod} is consistent with the mass-normalized modal basis commonly used in the structural engineering community.  Moreover, we observe that choosing the reduced mass matrix as $\Mhat=\Ir$ (without employing the reduced kinetic energy term $\widehat{T}_{\text{NN}}(\qhatdot)$ in step 2) yields accurate ROMs with {\Ra{fewer}} network parameters compared to the most expressive LOpInf-SpML ROM form~\eqref{eq:rom_form_general}. For this numerical example, we use four hidden layers with $\{64, 30, 20, 6\}$ units in the neural network parametrization for the reduced potential energy function $\widehat{U}_{\text{NN}}(\qhat)$ which leads to LOpInf-SpML ROMs with network parameters $N_{\text{param}}=2903$ for $r=4$, $N_{\text{param}}= 3031$ for $r=6$, and $N_{\text{param}}=3159$ for $r=8$.
\begin{table}
\caption{ {\Rb{Conservative rod model (relative state error comparison between four different methods)}}. Bold font highlights the lowest values in each column, and the strikethrough indicates that the POD-SpML approach failed to learn a stable ROM in $10^4$ epochs for $r=4$ and $r=8$.}
\begin{tabular}{l*{6}{c}r}
\toprule
& & Training data  & & & Test data &\\
\cmidrule(rl){2-4} \cmidrule(rl){5-7} 
    Method          & $r=4$ & $r=6$ & $r=8$  & $r=4$ & $r=6$ & $r=8$ \\
\hline
Intrusive POD ROM & $3.0 \times 10^{-2}$ & $1.3 \times 10^{-2}$ & $1.1 \times 10^{-2} $ & $8.6 \times 10^{-2}$ & $3.5 \times 10^{-2}$ & $2.6 \times 10^{-2} $ \\
LOpInf  ROM           & $3.1 \times 10^{-2}$ & $3.1 \times 10^{-2}$ & $3.1 \times 10^{-2}$ & $8.6 \times 10^{-2}$ & $8.6 \times 10^{-2}$ & $8.6 \times 10^{-2}$  \\
LOpInf-SpML  ROM          & $\mathbf{2.5 \times 10^{-2}}$ & $\mathbf{1.1 \times 10^{-2}}$ & $\mathbf{8.1 \times 10^{-3}} $  & $\mathbf{7.9 \times 10^{-2}}$ & $\mathbf{2.8 \times 10^{-2}}$ & $\mathbf{2.6 \times 10^{-2}} $  \\
POD-SpML ROM          & -----------  & $3.7 \times 10^{-2}$ &----------- & -----------  & $1.9 \times 10^{-1}$ &----------- \\
\bottomrule
\end{tabular}
\label{table}
\end{table}
\begin{figure}[h]
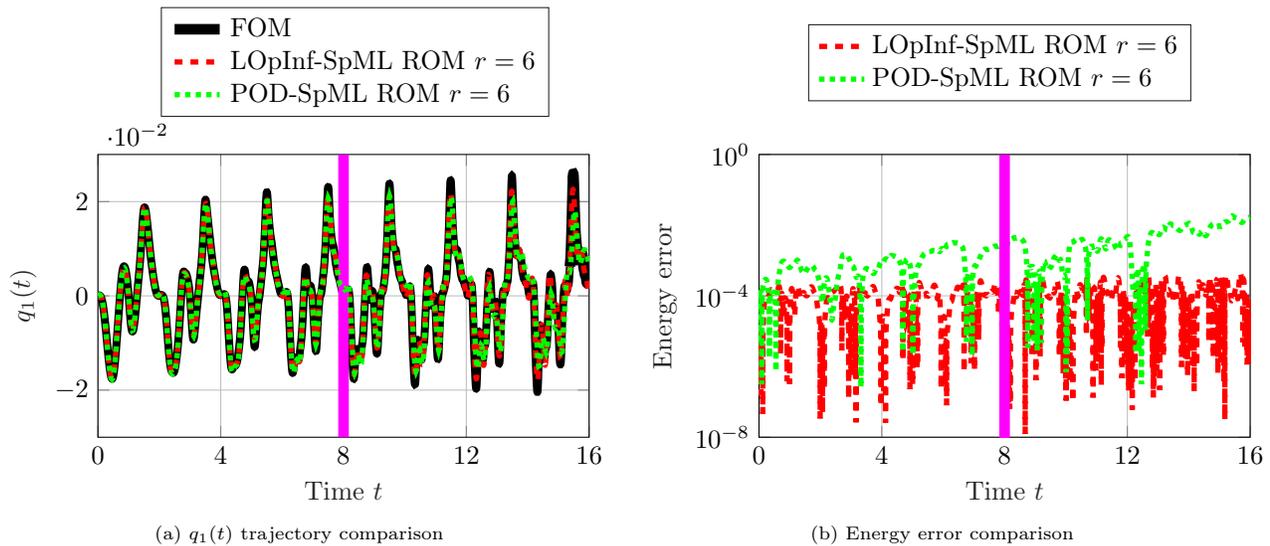

    \begin{subfigure}[b]{0.45\textwidth}
               \setlength\fheight{4.5 cm}
        \setlength\fwidth{\textwidth}
\input{Figures/conservative/q1_comp.tex}
        \subcaption[]{ $q_1(t)$ trajectory comparison}
        \label{fig:cons_q1}
    \end{subfigure}
    \hspace{0.4 cm}
        \begin{subfigure}[b]{0.45\textwidth}
                           \setlength\fheight{4.5 cm}
        \setlength\fwidth{\textwidth}
\input{Figures/conservative/energy_comp.tex}
       \subcaption[]{Energy error comparison}
         \label{fig:cons_energy}
    \end{subfigure}
    \caption{Conservative rod model. (a) The LOpInf-SpML ROM correctly predicts $q_1(t)$ trajectory $100\%$ outside the training time interval whereas the POD-SpML ROM provides inaccurate predictions after $t=12$. (b) The LOpInf- SpML ROM exhibits bounded energy error behavior with an energy error of approximately $10^{-4}$. The energy error for the POD-SpML ROM, on the other hand, slowly grows with time. The magenta line in both plots indicates the end of the training time interval at $T_{\text{train}}=8$.}
     \label{fig:cons}
\end{figure}
\begin{figure}[h]
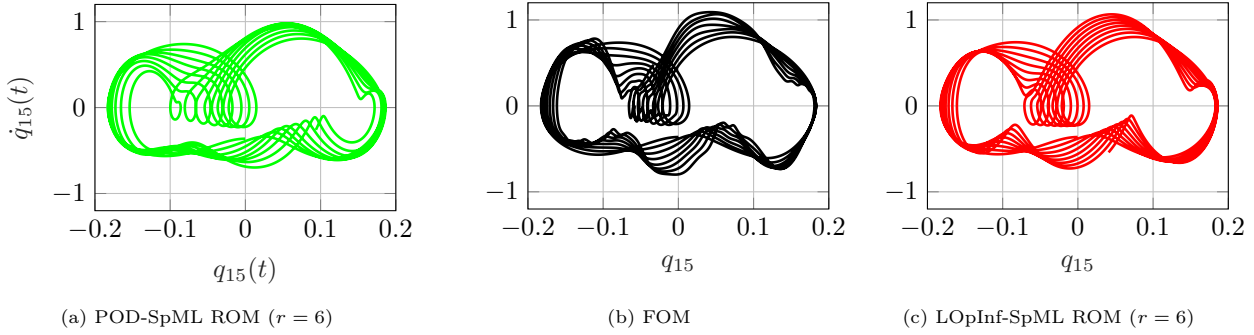

        \begin{subfigure}[b]{0.3\textwidth}
               \setlength\fheight{4 cm}
        \setlength\fwidth{\textwidth}
        \input{Figures/conservative/pp_spml_2.tex}
       \subcaption[]{POD-SpML ROM ($r=6$)}
                \label{fig:cons_pp_nnet}

    \end{subfigure}
    \hspace{0.5cm}
    \begin{subfigure}[b]{0.3\textwidth}
                   \setlength\fheight{4 cm}
        \setlength\fwidth{\textwidth}
\input{Figures/conservative/pp_fom_2.tex}
        \subcaption[]{FOM}
        \label{fig:cons_pp_fom}
    \end{subfigure}
    \hspace{-0.2cm}
        \begin{subfigure}[b]{0.3\textwidth}
                           \setlength\fheight{4 cm}
        \setlength\fwidth{\textwidth}
\input{Figures/conservative/pp_nnet.tex}
       \subcaption[]{LOpInf-SpML ROM ($r=6$)}
         \label{fig:cons_pp_spml}
    \end{subfigure}
    \caption{Conservative rod model. An accurate phase space portrait obtained using the LOpInf-SpML ROM demonstrates that the proposed approach has learned the underlying nonlinear dynamics whereas the POD-SpML ROM fails to capture the qualitative behavior of the FOM in phase space. }
     \label{fig:cons_pp}
\end{figure}
\begin{figure}[h]
    \centering
               \setlength\fheight{8.5 cm}
        \setlength\fwidth{\textwidth}
\input{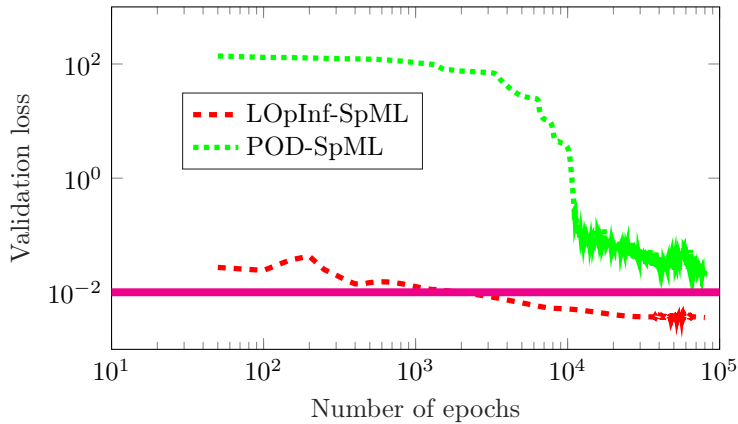}
    \caption{Conservative rod model. The LOpInf-SpML ROM based on the two-step learning approach starts at a significantly lower validation loss value and achieves the lowest error achieved by the POD-SpML ROM in $100\times$ {\Ra{fewer}} epochs. The horizontal magenta line indicates the lowest validation error achieved by the POD-SpML ROM.}
    \label{fig:cons_training}
\end{figure}
\begin{figure}[h]
    \centering
               \setlength\fheight{8.5 cm}
        \setlength\fwidth{\textwidth}
\input{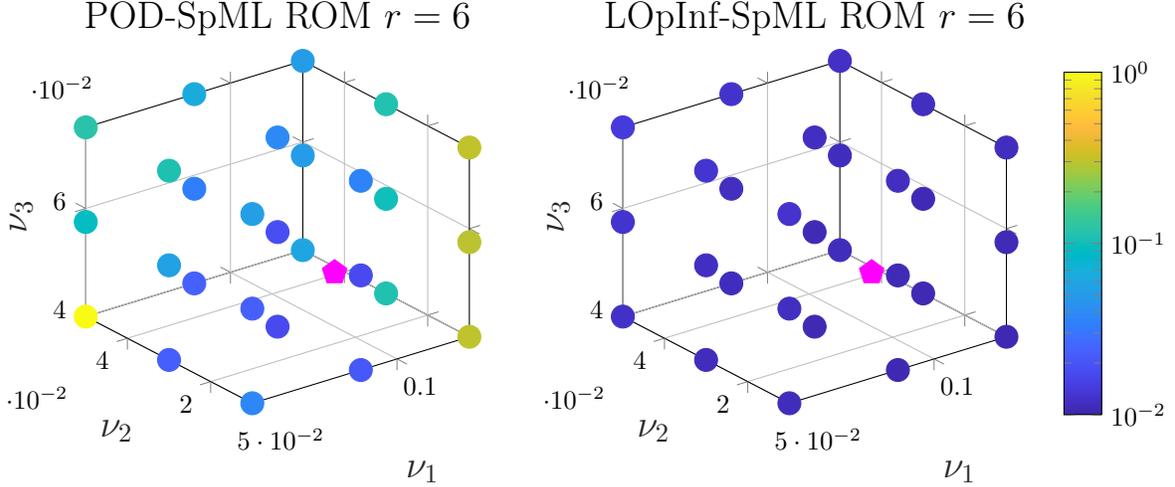}
    \caption{Conservative rod model. The LOpInf-SpML ROM is robust to perturbations in the training initial condition and achieves a relative state error of approximately $10^{-2}$ for unseen initial conditions whereas the POD-SpML ROM fails to provide accurate predictions for unseen initial conditions that are marginally different than the training initial condition. The magenta star in both plots represents the training initial condition.}
     \label{fig:cons_generalization}
\end{figure}
\subsubsection{Numerical results}
Table~\ref{table} compares the numerical performance of four different model reduction approaches for the conservative rod model example. The intrusive POD approach in this comparison derives nonlinear Lagrangian ROMs via the intrusive Galerkin projection of FOM operators onto the POD basis. The LOpInf method learns the best-fit linear Lagrangian ROMs nonintrusively from the data whereas the LOpInf-SpML approach combines the linear LOpInf ROM with structure-preserving neural networks to learn nonlinear Lagrangian ROMs in a nonintrusive manner. We also compare to a POD-SpML approach, which learns nonlinear Lagrangian ROMs by training the structure-preserving neural networks (without any prior) directly on the reduced snapshot data (projected onto the POD modes). The relative state error values reported in Table~\ref{table} compute
\begin{equation*}
        \text{relative state error}=\frac{\lVert \Q-\V\tilde{\Q}\rVert_F}{\lVert \Q\rVert_F},
    \end{equation*}
where $\tilde{\Q}$ is reduced snapshot data obtained by numerically integrating the learned ROM. For both training and test regimes, we observe that the relative state error decreases with an increase in the reduced dimension for the intrusive POD and the proposed LOpInf-SpML approach{\Rb{ with the LOpInf-SpML approach performing marginally better. This higher accuracy of the LOpInf-SpML approach, compared to the intrusive POD ROMs, can be explained through the lens of data-driven closure modeling, see~\cite{ahmed2021closures} and the references therein. In projection-based model reduction, data-driven closure modeling seeks to model the effect of the discarded POD modes on the ROM dynamics. The step 2 in the proposed LOpInf-SpML approach may be viewed as learning data-driven closure terms that capture the unresolved dynamics implicitly through the expressive polynomial-augmented MLPs employed in~\eqref{eq:Unn} and~\eqref{eq:Fnn}, and as a result, yielding LOpInf-SpML ROMs with lower relative state errors than the intrusive POD ROMs.}} The errors for the nonintrusive linear Lagrangian ROM based on LOpInf do not show improvement{\Rb{ with an increase in the reduced dimension. The state error leveling-off for LOpInf is due to two separate phenomena. First and foremost, this stagnation demonstrates the limitations of linear data-driven Lagrangian ROMs in capturing dynamic behavior that is inherently nonlinear. Second, theoretical analysis in~\cite{peherstorfer2020sampling} has shown that this state error leveling-off would occur even for linear FOMs because the projected snapshot data in~\eqref{eq:qhat} corresponds to non-Markovian dynamics in the reduced setting even though the FOM dynamics are Markovian}}\footnote{ {\Rb{The state error leveling-off for operator inference methods has been resolved by a re-projection sampling scheme in~\cite{peherstorfer2020sampling} for fully discrete FOMs with explicit time integrators. However, this scheme in its current form cannot be used for the mechanical FOMs considered in this work because they require implicit integrators for accurate numerical simulation.}}}. The POD-SpML approach yields the least accurate ROM in both training and test regimes, which shows that directly training a structure-preserving neural network on the reduced data is not a good strategy. Overall, the proposed LOpInf-SpML approach yields the most accurate ROMs in both training and test regimes for all three $r$ values.

The $q_1(t)$ trajectory plots in Figure~\ref{fig:cons_q1} compare the approximate ROM solutions against the benchmark FOM solution from $t=0$ to $T=16$. Both the POD-SpML ROM and the LOpInf-SpML ROM solutions are indistinguishable from the FOM solution in the training regime. In the test regime, we observe that the LOpInf-SpML ROM solution accurately tracks the FOM solution whereas the POD-SpML ROM solution is less accurate in capturing amplitude and  frequency content. In Figure~\ref{fig:cons_energy}, we observe that the proposed approach yields a structure-preserving ROM that exhibits bounded energy error behavior whereas the energy error for the POD-SpML ROM grows with time.

Phase space portraits are an invaluable tool for studying conservative dynamical systems. Figure~\ref{fig:cons_pp} compares the accuracy of the phase-space portraits in the $(q_{15},\dot{q}_{15})$ phase space. The phase space plot based on the LOpInf-SpML ROM in Figure~\ref{fig:cons_pp_spml} accurately captures the qualitative behavior of the phase space plot based on the FOM in Figure~\ref{fig:cons_pp_fom} whereas the phase space plot based on the POD-SpML ROM in Figure~\ref{fig:cons_pp_nnet} deviates away from the phase space plot based on the FOM. This comparison demonstrates that the proposed LOpInf-SpML approach has learned the nonlinear characteristics of the problem. {\Ra{Similarly}} to this representative example, the LOpInf-SpML approach demonstrates a higher accuracy than the POD-SpML approach in the phase-space portraits for other mass elements.

Figure~\ref{fig:cons_training} illustrates the decay in the validation loss for both approaches during the offline training stage. Both neural networks are trained from the reduced data using the same SpML architecture. Due to its two-step learning strategy, the proposed LOpInf-SpML approach starts at a validation loss value four orders of magnitude lower than the POD-SpML method. The validation loss for the LOpInf-SpML approach continues to decrease as training proceeds, remaining significantly lower than the POD-SpML ROM. {\Ra{The horizontal magenta line indicates the lowest validation error achieved by the POD-SpML approach. We observe that this error is achieved by the LOpInf-SpML approach with $100\times$ fewer epochs.}} This study highlights that using the LOpInf ROM for learning the linear ROM components provides higher accuracy along with significant computational savings in offline training.

Finally, we study the generalizability of the learned models by evaluating their accuracy for unseen initial conditions based on $(\nu_1,\nu_2,\nu_3)$ values that are different from the values considered in~\eqref{eq:ic}. We first simulate the FOM until $T=16$ for $27$ test initial conditions based on different combinations of $\nu_1\in\{5.0 \times 10^{-2}, 8.75 \times 10^{-2}, 1.25 \times 10^{-1} \}$, $\nu_2\in\{1.0 \times 10^{-2}, 3.0 \times 10^{-2}, 5.0 \times 10^{-2} \}$, and $\nu_3\in\{4.0 \times 10^{-2}, 5.75 \times 10^{-2}, 1.25 \times 10^{-1} \}$. We then simulate both LOpInf-SpML and POD-SpML ROMs for these $27$ test initial conditions and compare the relative state error values for each of those test initial conditions. The comparison in Figure~\ref{fig:cons_generalization} shows that the proposed approach achieves a relative state error of approximately $10^{-2}$ over a wide range of initial conditions, an order of magnitude more accurate than the POD-SpML approach.

 \subsection{Two-dimensional nonlinear membrane}
\label{sec: membrane}
We consider a two-dimensional nonlinear membrane model with internal damping to study the performance of the proposed approach for nonconservative mechanical systems. 
\subsubsection{FOM implementation}
We study a two-dimensional membrane with length $l_x=\sqrt{2}$ in the $x-$direction and length $l_y= 1/\sqrt{2}$ in the $y-$direction. The FOM for this numerical example is obtained using two-dimensional membrane elements with $N_x=21$ nodes in the $x-$direction and $N_y=13$ nodes in the $y-$direction. In these membrane elements, the in-plane stiffness comes from a pre-tension force. Each node in the two-dimensional membrane element has an out-of-plane displacement, which after applying the clamped boundary conditions leads to a FOM with $n = 240$ degrees of freedom.  

The linear stiffness associated with each element is defined via stiffness coefficient $k_x=n \cdot (l_x/N_x)^2$ in the $x-$direction and stiffness coefficient $k_y=n \cdot (l_y/N_y)^2$ in the $y-$direction. The corresponding linear stiffness matrix $\mathbf{K}$ is obtained by assembling elements based on their connectivity. The linear damping matrix is proportional to the stiffness matrix, so $\mathbf{C} = 10^{-4}\cdot\mathbf{K}$. The nonlinear behavior is modeled using cubic nonlinearities with nonlinear stiffness and nonlinear damping coefficients $k_{nl} = c_{nl} = 0.02n^{3/2} = 74.361$. Such cubic nonlinearities are commonly used for modeling complex nonlinearities exhibiting softening or hardening behavior, such as geometric nonlinearities~\cite{Kerschen2006PastPA}. The governing equations for the nonconservative FOM are
\begin{equation}
        \M\ddot{\q} + \C\qdot+  \frac{\partial \mathcal F_{\text{nl}} (\qdot)}{\partial \qdot} + \K\q +\frac{\partial U_{\text{nl}} (\q)}{\partial \q}=\bzero.  
          \label{eq:membrane_fom}
\end{equation}
{\Ra{Similarly}} to the conservative rod example, the proposed LOpInf-SpML approach learns a structure-preserving nonlinear ROM purely from the data, without any prior knowledge about the FOM~\eqref{eq:membrane_fom} for the two-dimensional nonlinear membrane. 
\subsubsection{Learning setup}
In this numerical example, we generate simulated data for the training and test datasets by integrating the nonconservative FOM~\eqref{eq:membrane_fom} with the Newmark integrator until the final time $T=35$. We use a fixed time step of $\dt=5 \times 10^{-3}$. We use data from $t=0$ to $t=15$ as the training data, data from $t=15$ to $t=17.5$ as the validation data, and data from $t=17.5$ to $T=35$ as the test data. {\Rb{We consider ROMs of size $r=10$ and $r=12$ to study the effect of an increase in the reduced dimension for the LOpInf ROMs, the POD-SpML ROMs, and the LOpInf-SpML ROMs. For comparison between the POD-SpML approach and the proposed LOpInf-SpML approach, we consider ROMs of size $r=12$ as the POD-SpML approach failed to learn a stable ROM in $10^4$ epochs for $r=10$.}} 

Due to the nonconservative nature of the problem, we consider a nonconservative nonlinear ROM of the form
\begin{equation}
            \qhatddot + \Chat\qhatdot + \frac{\partial \widehat{\mathcal F}_{\text{NN}}(\qhatdot)}{\partial \qhatdot} + \Khat\qhat + \frac{\partial \widehat{U}_{\text{NN}}(\qhat)}{\partial \qhat} = \bzero,
\end{equation}
where $\Khat$ and $\Chat$ are the linear reduced operators, and $\widehat{\mathcal F}_{\text{NN}}(\qhatdot)$ and $\widehat{U}_{\text{NN}}(\qhat)$ are the neural network parametrizations for the nonlinear dissipation function and the nonlinear potential energy function, respectively. {\Ra{Similarly}} to the conservative rod example in Section~\ref{sec: rod}, we observe that choosing the reduced mass matrix as $\Mhat=\Ir$ yields accurate and stable Lagrangian ROMs with {\Ra{fewer}} training parameters. Both $\widehat{\mathcal{F}}_{\text{NN}}(\qhatdot)$ and $\widehat{U}_{\text{NN}}(\qhat)$ are trained using a five-layer neural network with $ \{128,64,30,20,12\} $ units which leads to an LOpInf-SpML ROM with $N_{\text{param}}=24750$ network parameters. 
\begin{figure}[h]
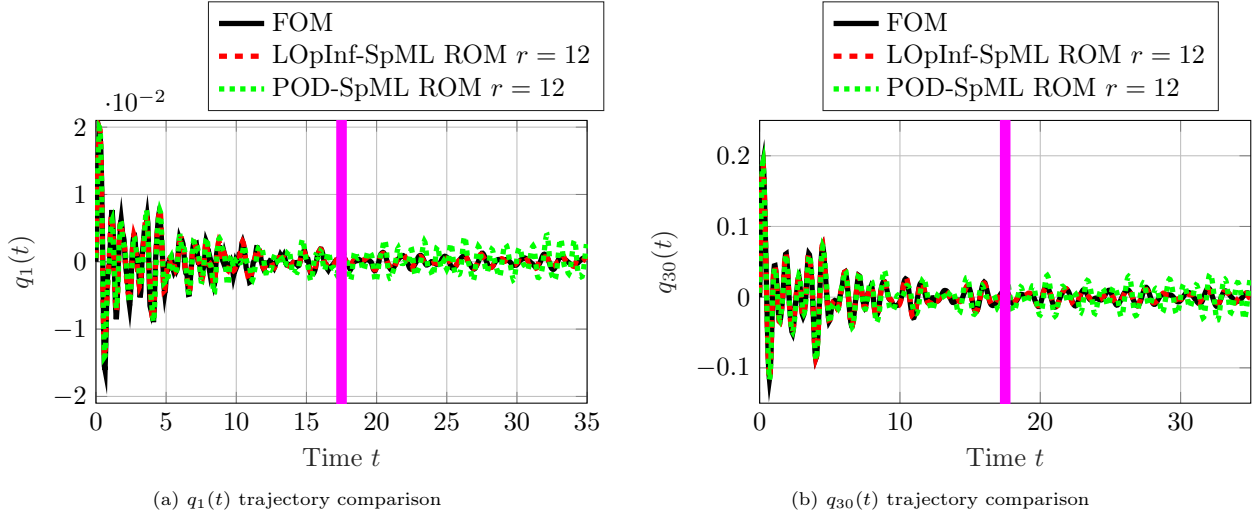

    \begin{subfigure}[b]{0.45\textwidth}
               \setlength\fheight{4.5 cm}
        \setlength\fwidth{\textwidth}
\input{Figures/membrane/q1_comp.tex}
        \subcaption[]{$q_{1}(t)$ trajectory comparison}
        \label{fig:membrane_traj1}
    \end{subfigure}
    \hspace{0.4 cm}
        \begin{subfigure}[b]{0.45\textwidth}
                           \setlength\fheight{4.5 cm}
        \setlength\fwidth{\textwidth}
\input{Figures/membrane/q30_comp.tex}
       \subcaption[]{$q_{30}(t)$ trajectory comparison}
         \label{fig:membrane_traj2}
    \end{subfigure}
    \caption{Nonconservative membrane model. Plots in (a) and (b) show that the LOpInf-SpML ROM of reduced dimension $r=12$ provides accurate predictions $100\%$ outside the training time interval whereas the POD-SpML ROM of size $r=12$ fails to learn the internal nonlinear damping which leads to trajectories with inaccurate amplitudes outside the training time interval. The magenta line indicates the end of the training time interval.}
     \label{fig:membrane_traj}
\end{figure}
\begin{figure}
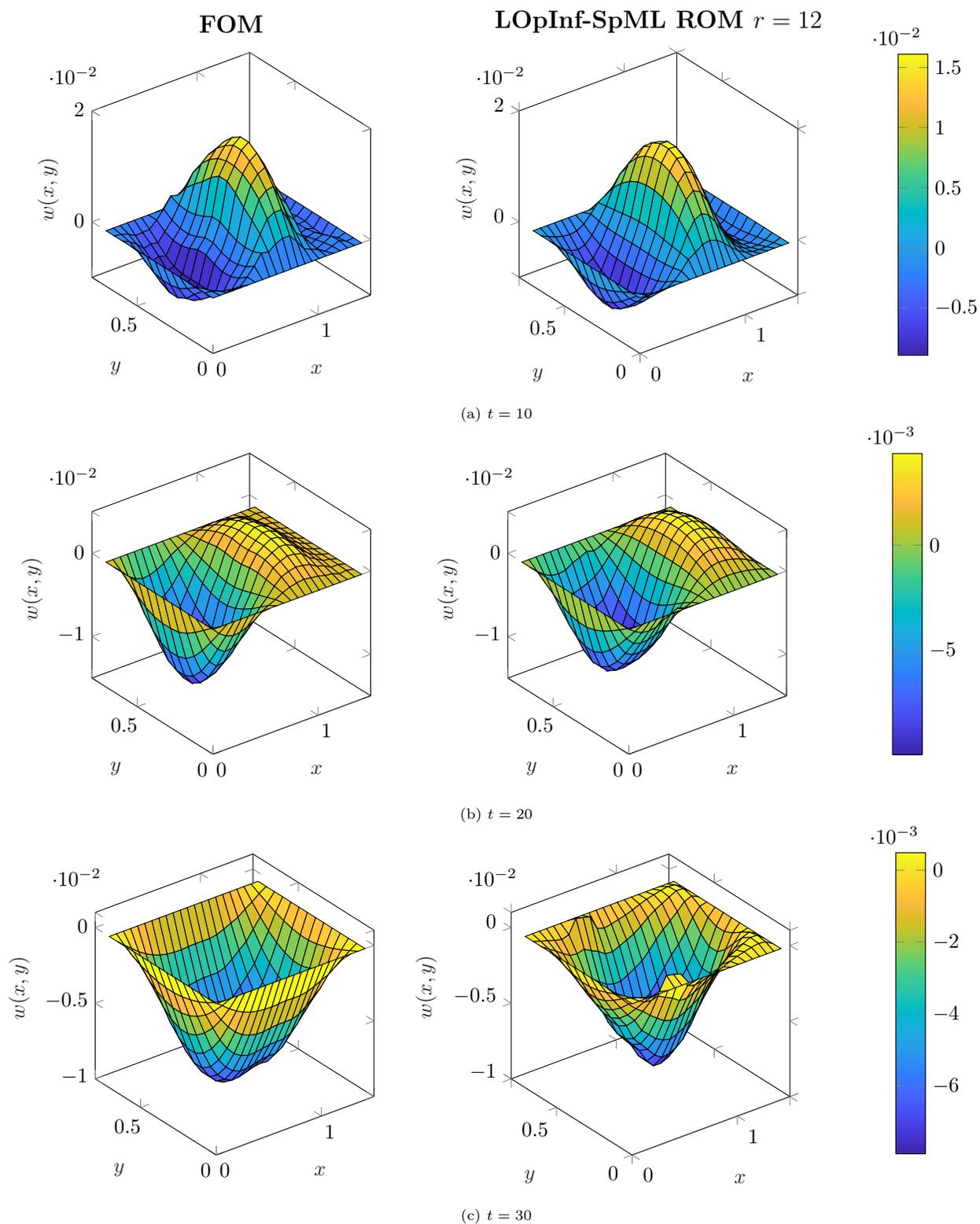

    \centering
    \begin{subfigure}[b]{0.95\textwidth}
       \setlength\fheight{9.5 cm}
        \setlength\fwidth{\textwidth}
\input{Figures/membrane/surf_10.tex}
        \subcaption[]{$t=10$}
         \label{fig:membrane_snapshot1}
    \end{subfigure}
    \begin{subfigure}[b]{0.95\textwidth}
     \setlength\fheight{9.5 cm}
        \setlength\fwidth{\textwidth}
\input{Figures/membrane/surf_20.tex}
       \subcaption[]{ $t=20$}
       \label{fig:membrane_snapshot2}
    \end{subfigure}
        \begin{subfigure}[b]{0.95\textwidth}
             \setlength\fheight{9.5 cm}
        \setlength\fwidth{\textwidth}
\input{Figures/membrane/surf_30.tex}
       \subcaption[]{$t=30$}
           \label{fig:membrane_snapshot3}
    \end{subfigure}
    \caption{Nonconservative membrane model. The membrane displacement comparison in plots (a), (b), and (c) shows that the LOpInf-SpML ROM provides accurate predictions over the entire computational domain.}
    \label{fig:membrane_snapshot}
 \end{figure}
\begin{figure}[h]
               \setlength\fheight{8.5 cm}
        \setlength\fwidth{\textwidth}
\input{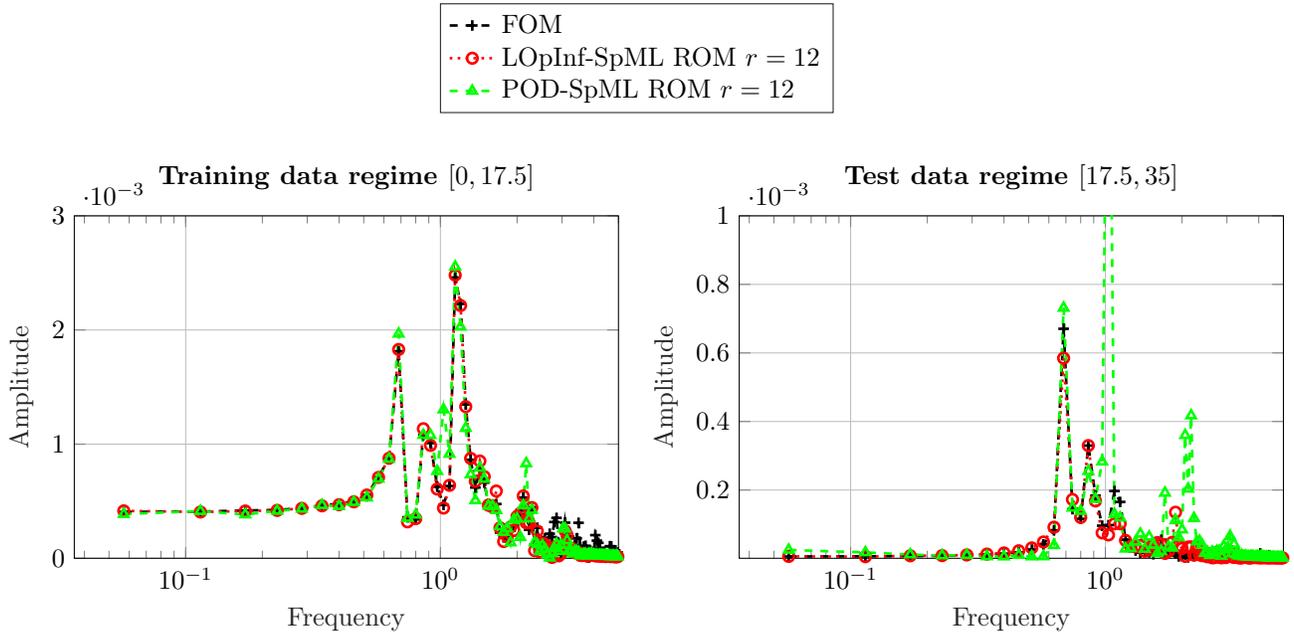}
    \caption{Nonconservative membrane model. In the training data regime, both ROMs capture the frequency content in the training data accurately. In the test regime, we observe that the LOpInf-SpML ROM predicts the frequencies in the system significantly better than the POD-SpML ROM.}
    \label{fig:membrane_freq}
\end{figure}
\subsubsection{Numerical results}
\begin{table}
\centering
\caption{ {\Rb{ Nonconservative membrane model (relative state error comparison between different nonintrusive methods). Bold font highlights the lowest values in each column, and the strikethrough indicates that the POD-SpML approach failed to learn a stable ROM in $10^4$ epochs for $r=10$.}}}
\begin{tabular}{l*{6}{c}r}
\toprule 
&& {\Rb{Training data}}  &  & {\Rb{Test data}}  &&\\
\cmidrule(rl){2-3} \cmidrule(rl){4-6} 
     {\Rb{Method }}          &  {\Rb{$r=10$}}  &  {\Rb{$r=12$}}  &  {\Rb{$r=10$}}   &  {\Rb{$r=12$}}   \\
\hline
 {\Rb{LOpInf  ROM}}            &  {\Rb{$2.6 \times 10^{-1}$}}  &  {\Rb{$2.6 \times 10^{-1}$}}  &  {\Rb{$9.7 \times 10^{-1}$}}  &  {\Rb{$9.7 \times 10^{-1}$ }}   \\
 {\Rb{LOpInf-SpML  ROM }}          &  {\Rb{$\mathbf{5.6 \times 10^{-2}}$}}  &  {\Rb{$\mathbf{5.1 \times 10^{-2}}$}}  &  {\Rb{$\mathbf{3.3 \times 10^{-1}} $}}   &  {\Rb{$\mathbf{1.9 \times 10^{-1}}$}}    \\
 {\Rb{POD-SpML ROM }}          &  {\Rb{-----------}}   &  {\Rb{$2.0 \times 10^{-1}$}}  & {\Rb{-----------}}  &  {\Rb{$1.0 \times 10^{0}$}} \\
\bottomrule
\end{tabular}
\label{table2}
\end{table}
{\Rb{Table~\ref{table2} compares the numerical performance of the three different nonintrusive model reduction approaches for the nonconservative membrane example.  For both training and test regimes, we observe that the relative state error for the LOpInf-SpML approach decreases as we increase the reduced dimension from $r=10$ to $r=12$ whereas the errors for the nonintrusive LOpInf ROMs do not show any improvement due to their linear ROM form. The POD-SpML ROM of dimension $r=12$ performs marginally better than the LOpInf ROM of dimension $r=12$ in the training data regime but fails to provide accurate predictions in the test data regime. Similarly to the conservative example in Section~\ref{sec: rod}, the proposed LOpInf-SpML approach yields the most accurate ROMs for the nonconservative membrane example in both training and test regimes for both $r$ values. }}

In Figure~\ref{fig:membrane_traj}, we compare the ROM solutions with the FOM solution for two different states. The $q_1(t)$ trajectory plots in Figure~\ref{fig:membrane_traj1} show that the LOpInf-SpML ROM captures the decay in the FOM solution accurately whereas the POD-SpML ROM provides inaccurate predictions outside the training data regime. For the $q_{30}(t)$ trajectory plots in Figure~\ref{fig:membrane_traj2}, we observe that the LOpInf-SpML ROM solution is indistinguishable from the FOM solution which demonstrates the proposed approach's ability to learn the internal nonlinear damping of the full-order model. The POD-SpML ROM solution, on the other hand, fails to learn the internal damping which leads to substantially higher errors in the test regime.

Figure~\ref{fig:membrane_snapshot} compares the FOM solution with the ROM solution at different time instances. In Figure~\ref{fig:membrane_snapshot1}, the LOpInf-SpML ROM provides an approximate full-field solution that agrees with the FOM solution. In Figure~\ref{fig:membrane_snapshot2}, we observe that the proposed approach accurately predicts the membrane displacements over the entire two-dimensional domain. Even though the solution at $t=30$ in Figure~\ref{fig:membrane_snapshot3} has been significantly damped out, the ROM solution predicts the displacement amplitude over the two-dimensional domain with reasonable accuracy.

Finally, we study the frequency characteristics of this nonconservative problem by analyzing the time-series data for the first node. We study the spectral information in the training data and the test data separately. All the models are numerically integrated with a time step of $\dt=5 \times 10^{-3}$, so the sampling frequency for this setting is $200~\si{\hertz}$. We apply the standard fast Fourier transform to obtain the frequency-amplitude curves in Figure~\ref{fig:membrane_freq}. We observe that both LOpInf-SpML and POD-SpML capture the frequency content in the training data regime accurately with the LOpInf-SpML ROM performing marginally better. In the test regime, we observe that the amplitude-frequency curve for the LOpInf-SpML ROM agrees with the FOM amplitude-frequency curve whereas the POD-SpML ROM only performs well for frequencies below $0.5~\si{\hertz}$. This spectral analysis shows that the ROM learned with the proposed LOpInf-SpML approach has learned the damped dynamics accurately.

\subsection{Half Brake-Reuß beam}
\label{sec: joint}
Accurate prediction of the dynamics of jointed structures remains a challenging problem due to the strong nonlinearities at the frictional interfaces found in joints. The main purpose of a joint in an engineering structure is to connect two separate substructures in a stiff manner. One of the principal roadblocks to accurate high-fidelity modeling is the lack of understanding of the physics behind the energy dissipation mechanisms at the joints. In part, this lack of understanding is due to the multiscale nature of interfacial mechanics for jointed structures where the influencing sources can range from macro-scale geometry and loading to nano-scale grain boundaries, see~\cite{brake2017mechanics} for more details. Due to the problem-specific and dynamic nature of interfacial mechanics, it is unrealistic to expect a universal law of friction that is predictive across most types of interfaces. For such problems, the structural and mechanical engineering community is mainly interested in developing a predictive model for different types of joints such as lap joints, dove-tail joints, fir-tree joints, and tape joints. In this section, we apply the proposed learning approach to learn a predictive model of a beam structure with a lap joint from a dataset consisting of experimental measurements. The dataset considered in this numerical example is obtained from the experimental setup in~\cite{CHEN2022108401, JIN2022108402} where the authors used high-speed cameras combined with digital image correlation to obtain the full-field response of the structure known as the "Half Brake-Reuß beam". The data was obtained from the authors' repository \footnote{https://github.com/mattiacenedese/BRBtesting}. 
\subsubsection{Experimental data collection using digital image correlation measurements}
\begin{figure}[h]%
\centering
\includegraphics[width=0.9\textwidth]{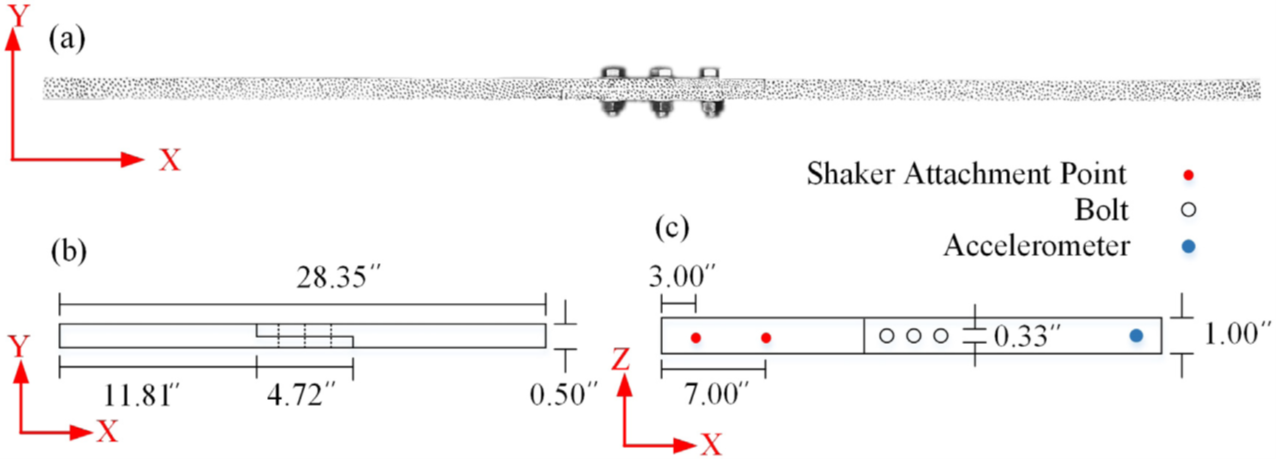}
  \caption{Half Brake-Reuß beam. Geometry of the jointed structure including a photo with digital image correlation speckle pattern shown (a), top view (b), and front view (c). Figure reprinted from~\cite{CHEN2022108401}, Copyright (2022), with permission from Elsevier and the authors.}
  \label{fig:hbrb}
\end{figure}

The half Brake-Reuß beam is a modification of the Brake--Reuß beam~\cite{brake2019observations}, which is a widely used benchmark structure for studying nonlinear dynamics in jointed systems. The half Brake-Reuß beam consists of two beams that are joined through a three-bolt lap joint as shown in Figure~\ref{fig:hbrb}. The lap joint is the most common type of interface found in built-up structures and is defined as the mating of two components through a bolted connection.

The free-response data considered in this study was obtained by first exciting the beam at a specific frequency of interest (i.e., at the natural frequency of the mode of interest) with a fixed sinusoidal signal and then detaching the shaker once resonance is identified. Even though this procedure is meant to isolate a mode of interest, the full-field response includes multiple modes due to modal coupling resulting from the nonlinearity. The nonlinearity in this structure arises due to energy dissipation caused by slipping and possible intermittent contact in the joint.

For the free-response experiments, the beam structure was suspended horizontally using bungee cords to simulate the free-free boundary conditions. Using the digital image correlation technique, the vibrational motion of $n=206$ measurement points was tracked for $T=3.06$~\si{\s} with a sampling frequency of $f_{\text{sampling}}=5000$~\si{\hertz}. The recorded images from this experiment were post-processed to build a dataset containing the time history of the displacements of the measurement points.
\subsubsection{Learning setup} 
For this example with experimental measurements, we use data from $t=0$~\si{\s} to $t=0.92$~\si{\s} as the training data, data from $t=0.92$~\si{\s} to $t=1.02$~\si{\s} as the validation data, and data from $t=1.02$~\si{\s} to $T=3.06$~\si{\s} as the test data. Joints introduce two qualitatively important features to a structure: amplitude-dependent stiffness and amplitude-dependent damping. The aim of this study is to use the LOpInf-SpML ROM as a surrogate model for the jointed beam structure and provide accurate predictions of these qualitative properties outside the training data regime\footnote{{\Rb{For this example, the POD-SpML ROM approach failed to learn stable ROMs in $10^4$ epochs from reduced dimension $r=3$ to $r=6$. The LOpInf ROMs, on the other hand, can not capture the nonlinear characteristics of the jointed structure due to their linear nature.}}}. To accurately capture the amplitude-dependent frequency and damping characteristics of the jointed structure, we consider the most expressive ROM form~\eqref{eq:rom_form_general} in this study. In addition to learning the neural network parametrizations for the dissipation and potential energy functions, we also learn a reduced kinetic energy term $\widehat{T}_{\text{NN}}(\qhatdot)$ that leads to a symmetric positive-definite reduced mass matrix $\Mhat_{\text{NN}}$ in the offline training phase, see~\eqref{eq:Mnn_new}. For this example, the reduced kinetic energy term $\widehat{T}_{\text{NN}}(\qhatdot)$ in~\eqref{eq:Tnn} is parametrized with nine unknown network parameters. Both $\widehat{\mathcal{F}}_{\text{NN}}(\qhatdot)$ and $\widehat{U}_{\text{NN}}(\qhat)$ are trained using a four-layer neural network with $ \{16,16,16,16\} $ units which leads to an LOpInf-SpML ROM with $N_{\text{param}}=2538$ network parameters. 
%
\begin{figure}[h]
    \centering
    \begin{subfigure}[b]{0.45\textwidth}
        \includegraphics[width=\textwidth, height=5.25 cm]{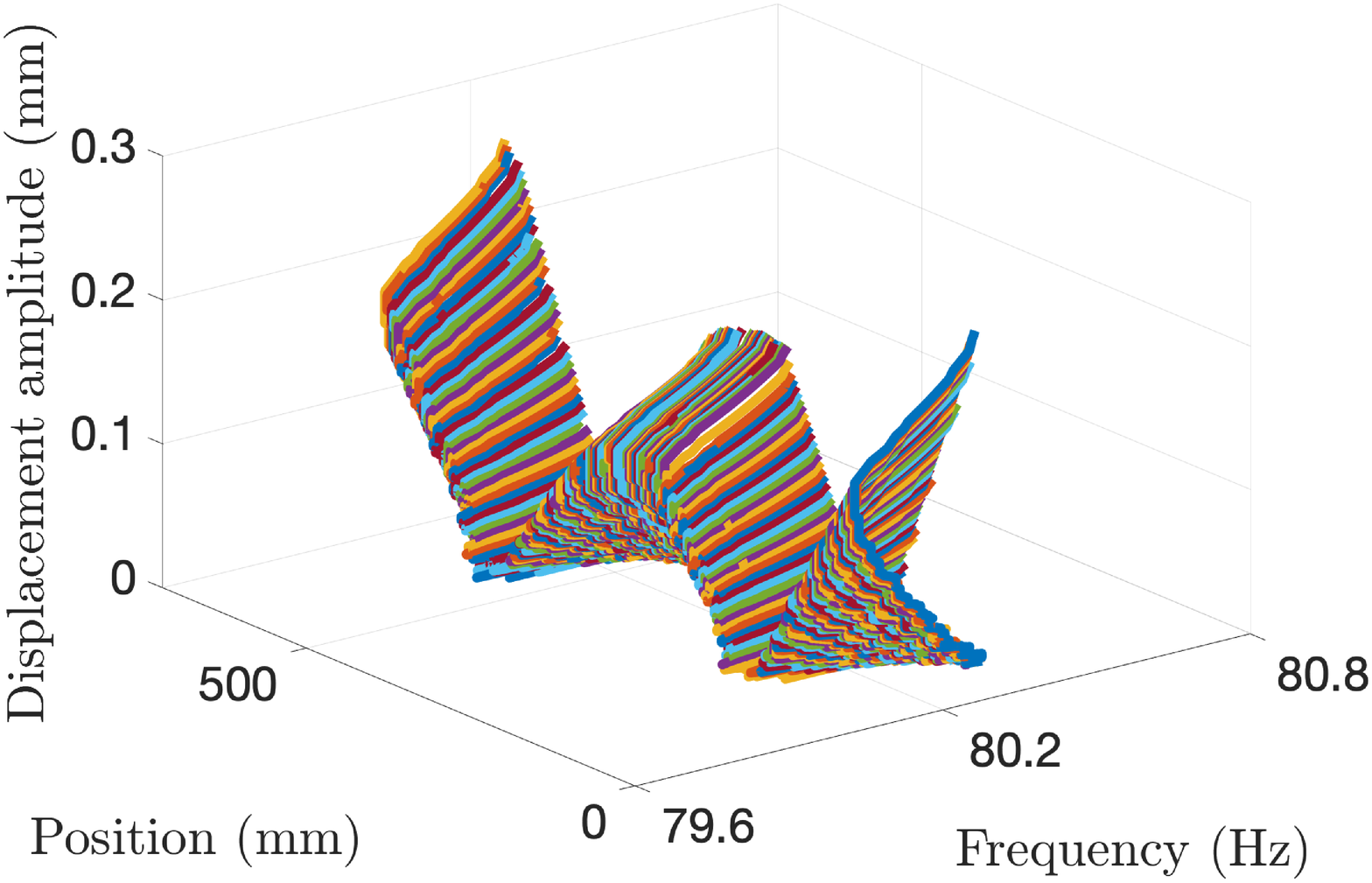}
        \subcaption[]{Experimental data}
        \label{figure:freq_dis_fom}
    \end{subfigure}
    \begin{subfigure}[b]{0.45\textwidth}
        \includegraphics[width=\textwidth, height=5.25 cm]{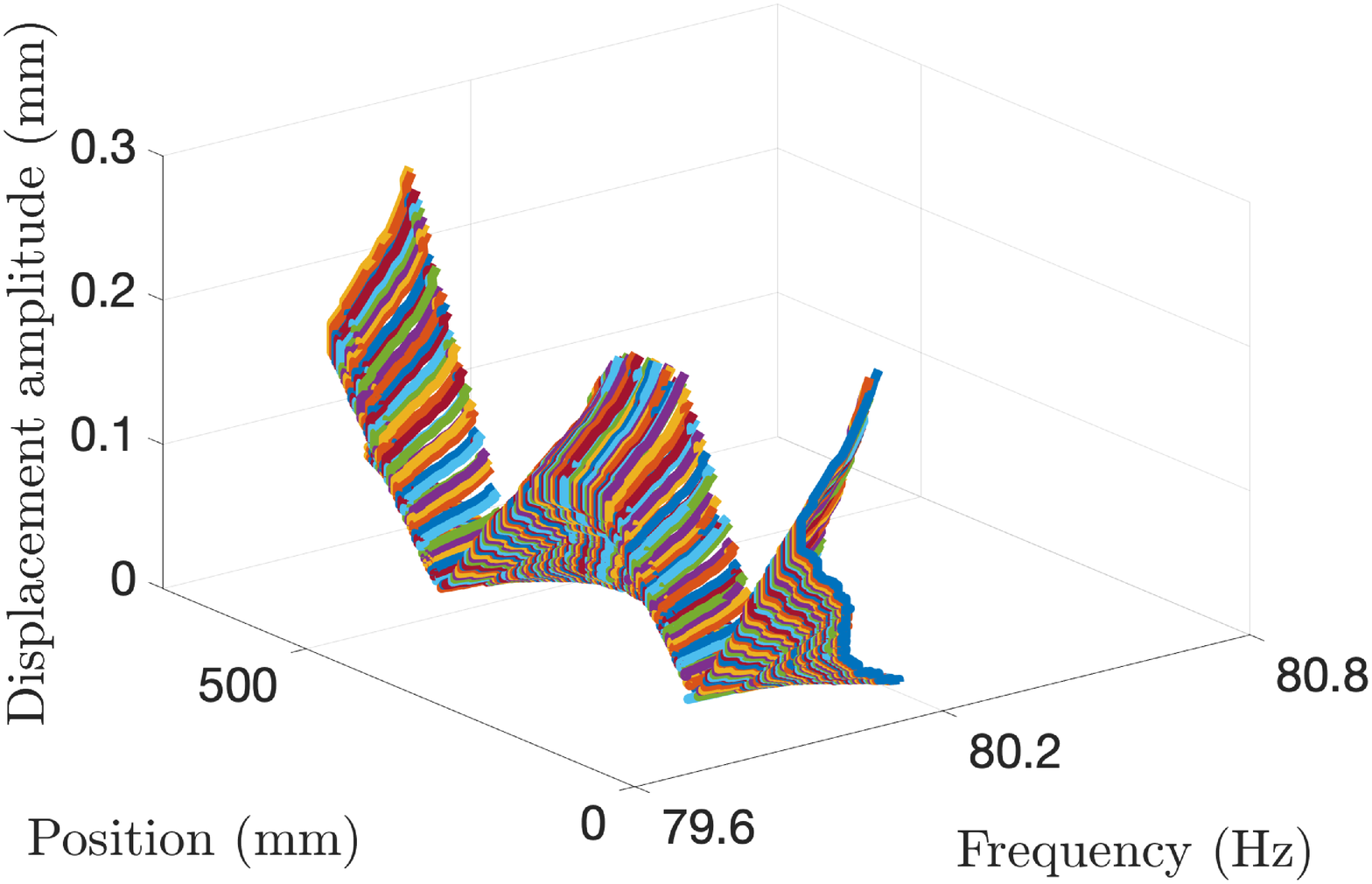}
       \subcaption[]{ LOpInf-SpML ROM $r=3$}
       \label{figure:freq_dis_rom}
    \end{subfigure}
    \caption{Half Brake-Reuß beam. The LOpInf-SpML ROM of size $r=3$ accurately predicts the amplitude-dependent frequency characteristics and yields backbone curves that appear to agree with the backbone curves obtained directly from the experimental data.}
    \label{figure:freq_dis}
\end{figure}
\begin{figure}[h]
    \centering
    \begin{subfigure}[b]{0.45\textwidth}
        \includegraphics[width=\textwidth, height=5.25 cm]{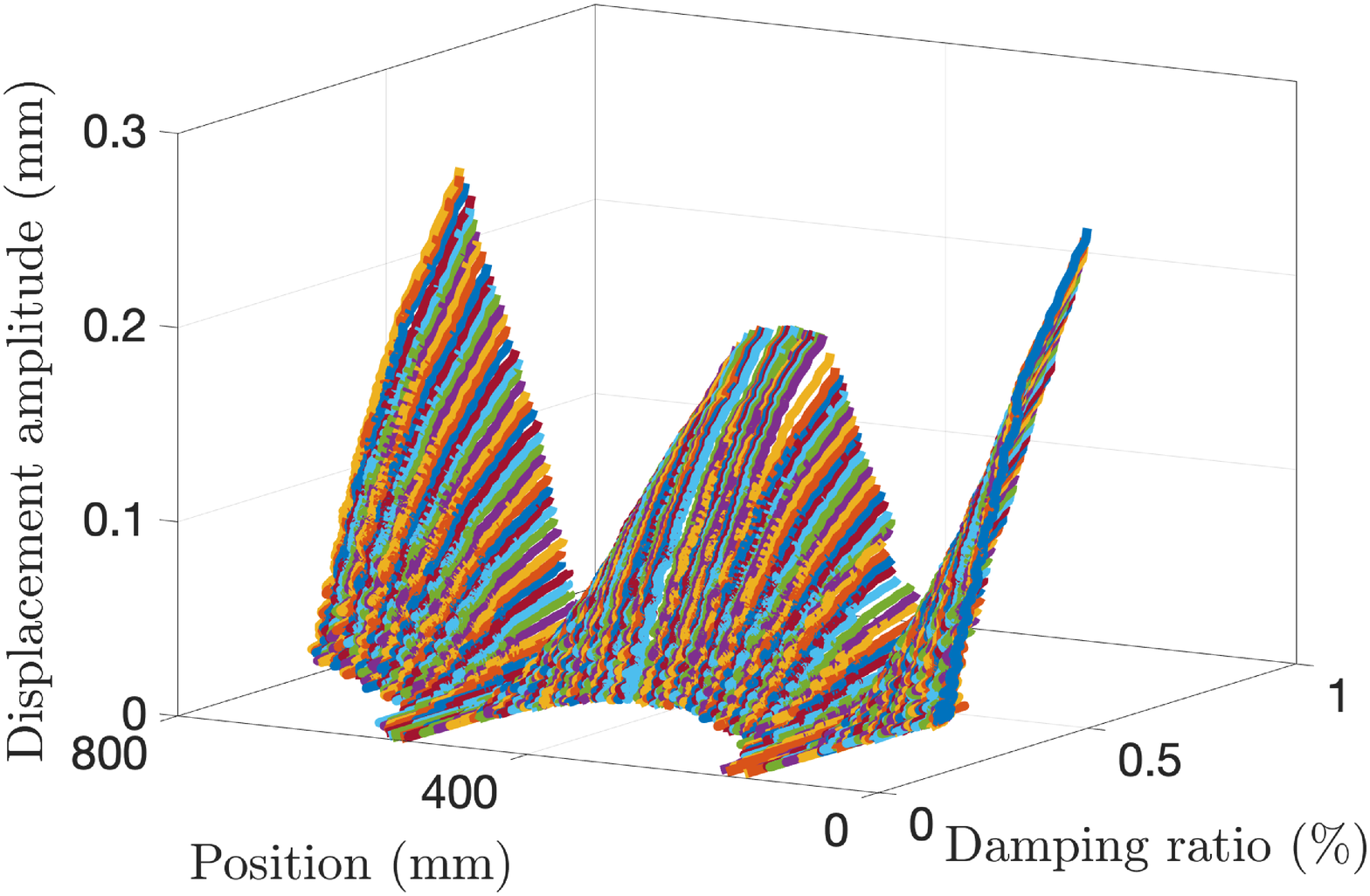}
        \subcaption[]{Experimental data}
        \label{figure:damping_dis_fom}
    \end{subfigure}
    \hspace{0.5cm}
    \begin{subfigure}[b]{0.45\textwidth}
        \includegraphics[width=\textwidth, height=5.25 cm]{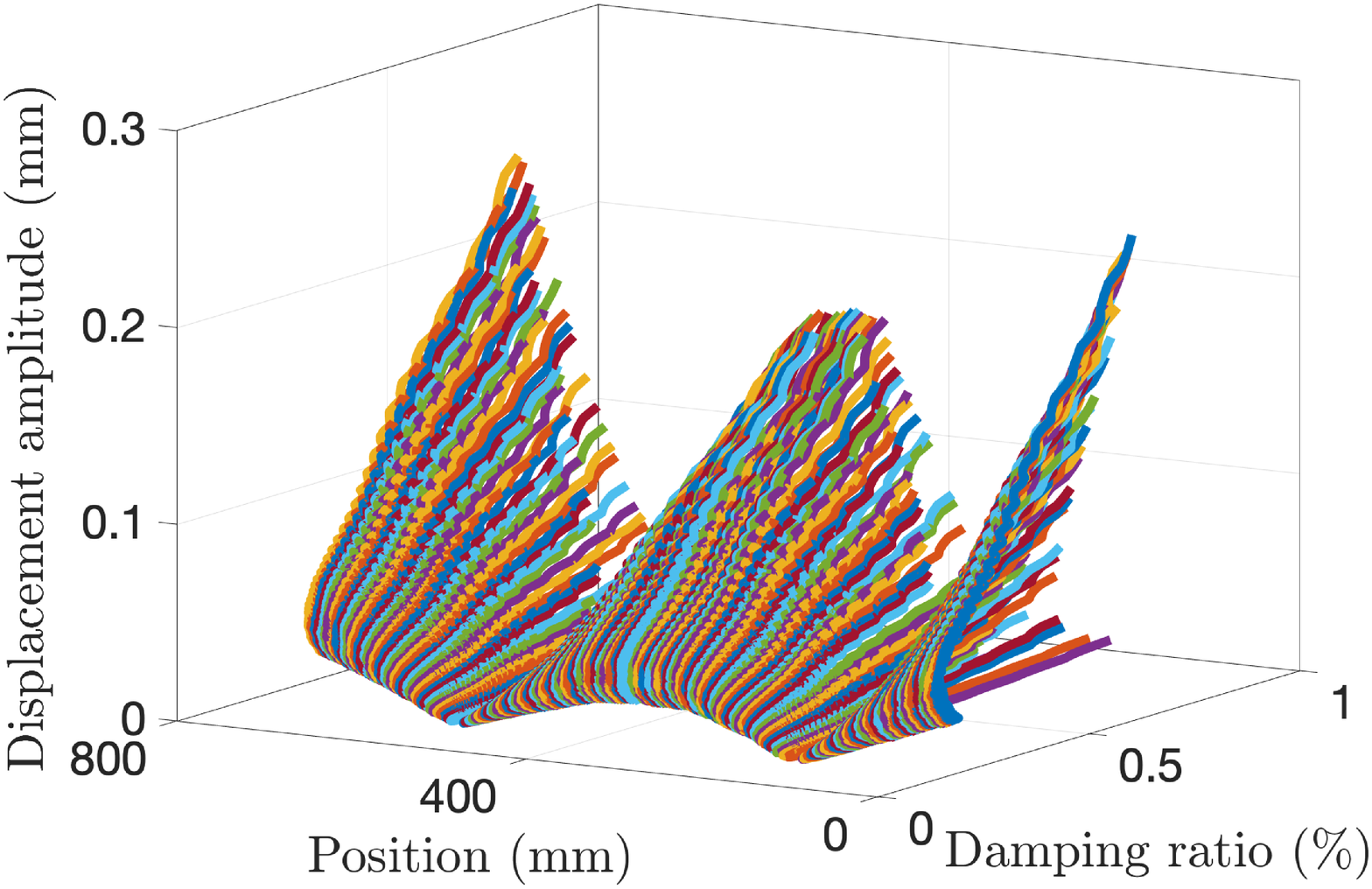}
       \subcaption[]{ LOpInf-SpML ROM $r=3$}
               \label{fig:dis_rom}
    \end{subfigure}
    \caption{Half Brake-Reuß beam. The amplitude-dependent damping plots based on the LOpInf-SpML ROM of dimension $r=3$ are reasonably similar to the plots obtained from the experimental data.}
     \label{figure:damping_dis}
\end{figure}
\begin{figure}[h]
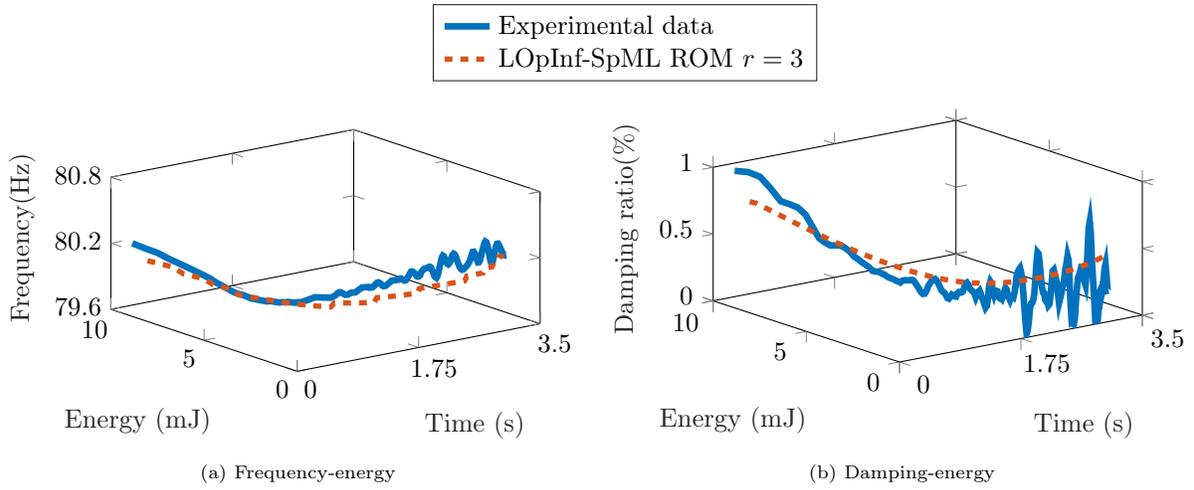

    \centering
    \begin{subfigure}[b]{0.45\textwidth}
    \setlength\fheight{6 cm}
        \setlength\fwidth{\textwidth}
\input{Figures/joint/freq_energy_comp.tex}
        \subcaption[]{Frequency-energy}
        \label{figure:energy_freq}
    \end{subfigure}
    \begin{subfigure}[b]{0.45\textwidth}
    \setlength\fheight{6 cm}
        \setlength\fwidth{\textwidth}
\input{Figures/joint/damping_energy_comp.tex}
       \subcaption[]{ Damping-energy}
       \label{figure:energy_damping}
    \end{subfigure}
    \caption{Half Brake-Reuß beam. The proposed approach accurately predicts the frequency-energy curve in (a) and captures the overall trend of the challenging damping-energy curve in (b).}
     \label{figure:energy_joint}
\end{figure}
\subsubsection{Numerical results}
Figure~\ref{figure:freq_dis_fom} shows the amplitude-dependent frequency curves of the jointed structure based on the experimental measurements at different points along the length of the beam. In Figure~\ref{figure:freq_dis_rom}, we observe that the LOpInf-SpML ROM accurately predicts the frequency range with less than $0.5\%$ error. More importantly, the amplitude-dependent frequency curves for the learned ROM has the same shape as the one obtained from the experimental data which demonstrates that the proposed approach has learned the nonlinear characteristics of the problem.

In Figure~\ref{figure:damping_dis}, we obtain the amplitude-dependent damping plots using the LOpInf-SpML ROM and compare them with the plots based on the experimental data. Even though the proposed ROM predicts damping ratio values higher than the ones based on the experimental data for some points, the amplitude-damping curves obtained using the ROM in Figure~\ref{fig:dis_rom} are reasonably similar to the plots obtained from the experimental data in Figure~\ref{figure:damping_dis_fom}.

As an alternative to displacement-based plots,  Figure~\ref{figure:energy_joint} shows the nonlinear characteristics of the jointed structure through energy-dependent plots. The total energy of the jointed structure consists of kinetic energy $T(\qdot)$ and potential energy $U(\q)$. The standard way to compute the total energy is to assume the potential energy to be zero at the equilibrium position and then calculate the kinetic energy at the equilibrium position
\begin{equation*}
        T(\qdot)=\sum_{i=1}^{n=206}\frac{1}{2}m\dot{q}_i^2,
\end{equation*}
where $\dot{q}_i$ is the velocity of the $i$th measurement point, and $m=M/n$ is the lumped mass of each measurement point. Since the energy calculation utilizes information about the motion at all the points across the length of the beam, the energy-dependent plots provide a more global perspective on the mechanical system under analysis. The frequency-energy plots in Figure~\ref{figure:energy_freq} show that the LOpInf-SpML ROM of reduced dimension $r=3$ accurately predicts the time-evolution of the frequency-energy relationship. In Figure~\ref{figure:energy_damping} we observe that the learned ROM accurately captures the overall energy-damping trend but predicts damping ratio values higher than the ones obtained from the experimental data. We note that the energy-dependent plots based on the experimental data are known to demonstrate increased variability towards small amplitudes due to the degradation of the signal-to-noise ratio in the digital image correlation measurements.
\section{Conclusions}
\label{sec:conclusions}
We have presented an ML-enhanced model reduction method that learns structure-preserving ROMs of nonlinear mechanical systems from data in a nonintrusive manner. The presented method parametrizes the ROM in terms of the mechanical problem's Lagrangian structure and then learns the reduced operators in a structure-preserving way to ensure that the learned operators are Lagrangian. The proposed learning algorithm uses a POD basis to project the high-dimensional data onto a low-dimensional subspace and then derives the Lagrangian ROM by (i) inferring the linear reduced operators using a structure-preserving nonintrusive operator inference method, and then (ii) learning nonlinear reduced operators using a structure-preserving machine learning method. The proposed LOpInf-SpML framework is well-suited for engineering applications where the data are generated from complicated high-fidelity computational models or from experimental measurements.

The numerical experiments for conservative and nonconservative mechanical systems and for simulated and experimental data demonstrate the wide applicability of the proposed approach. The nonlinear conservative rod model shows that the LOpInf-SpML approach learns long-time stable ROMs with bounded energy error, while also providing accurate predictions for unseen initial conditions. Compared to the POD-SpML approach, the proposed LOpInf-SpML approach achieves significant computational savings in the offline training phase which highlights the key role played by the linear Lagrangian ROM in our two-step learning algorithm. The two-dimensional nonlinear membrane example demonstrates the proposed method's ability to learn accurate ROMs for nonconservative mechanical systems with nonlinear damping. For the half Brake-Reuß Beam example, the learned ROM accurately captures the amplitude-dependent frequency and damping characteristics of the jointed structure from a dataset consisting of digital image correlation measurements. This example highlights the ability of the proposed method to learn a computationally efficient predictive model from experimental observations for applications where the underlying physics is not well understood. 

This work has motivated a number of future research directions. The proposed method uses linear basis approximations, which can make offline training prohibitively expensive if the problem requires a large basis size. For such problems, the proposed approach could be combined with autoencoders~\cite{lee2020model,buchfink2023symplectic} or structure-preserving nonlinear manifolds~\cite{sharma2023symplectic} to reduce the computational costs in the offline phase. In another direction, the proposed LOpInf-SpML approach could be extended to mechanical problems with geometric nonlinearities. Finally, we would like to apply our framework to learn models from noisy and sparse data.
\section*{Acknowledgments}
We thank Prof. Matthew Brake of Rice University (Houston, Texas) for sharing the experimental dataset for the jointed structure numerical example in Section~\ref{sec: joint}. H. Sharma and B. Kramer were in part financially supported by the Ministry of Trade, Industry and Energy (MOTIE) and the Korea Institute for Advancement of Technology (KIAT) through the International Cooperative R\&D program (No.~P0019804, Digital twin based intelligent unmanned facility inspection solutions) and the Applied and Computational Analysis Program of the Office of Naval Research under award N000142212624. D. Najera-Flores and M. Todd were funded by Sandia National Laboratories. Sandia National Laboratories is a multi-mission laboratory managed and operated by National Technology and Engineering Solutions of Sandia, LLC., a wholly owned subsidiary of Honeywell International, Inc., for the U.S. Department of Energy’s National Nuclear Security Administration under contract DE-NA-0003525. The University of California San Diego acknowledges subcontract Agreement 2,169,310 from Sandia National Laboratories for its participation in this work.
\bibliographystyle{vancouver}
\bibliography{main}
\end{document}